\documentclass[usenatbib,fleqn]{mnras}

\let\jnlstyle=\rm
\def\refjnl#1{{\jnlstyle#1}}

\def\apj{\refjnl{ApJ}}                 % Astrophysical Journal
\def\apjl{\refjnl{ApJ}}                % Astrophysical Journal, Letters
\def\apjs{\refjnl{ApJS}}               % Astrophysical Journal, Supplement
\def\ao{\refjnl{Appl.~Opt.}}           % Applied Optics
\def\aap{\refjnl{A\&A}}                % Astronomy and Astrophysics
\def\prd{\refjnl{Phys.~Rev.~D}}        % Physical Review D
\font \bolditalics = cmmib10
\def\bx#1{\leavevmode\thinspace\hbox{\vrule\vtop{\vbox{\hrule\kern1pt
        \hbox{\vphantom{\tt/}\thinspace{\bf#1}\thinspace}}
      \kern1pt\hrule}\vrule}\thinspace}
\def \vc #1{{\textfont1=\bolditalics \hbox{$\bf#1$}}}

\def\thetavc{{\vc \theta}}

\input{psfig.sty}
\usepackage{graphicx}
\usepackage{epsfig}
\usepackage{amssymb}
\usepackage{amsmath}
\usepackage{amsfonts}
\usepackage{txfonts}
\usepackage{multirow}
\usepackage{afterpage}
\usepackage{color}

\usepackage{hyperref}
\def\be{\begin{equation}}
\def\ee{\end{equation}}
\def\ba{\begin{eqnarray}}
\def\ea{\end{eqnarray}}

\def \msun{{\rm M}_{\odot}}

\def\der{{\rm d}}

\def\msun{\textrm{M}_{\odot}}

\title[RCSLenS: Cross-correlation with tSZ]{Cross-correlating \textit{Planck} tSZ with RCSLenS weak lensing: \\ Implications for cosmology and AGN feedback}
\author[Hojjati, Tr\"oster, Harnois-Deraps, McCarthy, Van Waerbeke et al.]{Alireza Hojjati$^{1}$\thanks{E-mail: ahojjati@phas.ubc.ca}, Tilman Tr\"oster$^{1}$\thanks{E-mail: troester@phas.ubc.ca}, Joachim Harnois-D\'{e}raps$^{2}$, Ian G. McCarthy$^{3}$,\newauthor   Ludovic van Waerbeke$^{1,4}$, Ami Choi$^{2}$, Thomas Erben$^{5}$, Catherine Heymans$^{2}$,\newauthor Hendrik Hildebrandt$^{5}$, Gary Hinshaw$^{1,4,6}$,Yin-Zhe Ma$^{7}$,  Lance Miller$^{8}$, Massimo Viola$^{9}$,  \newauthor  Hideki Tanimura$^{1}$ 
\\
$^{1}$Department of Physics and Astronomy, University of British Columbia, Vancouver, V6T 1Z1, BC, Canada\\
$^{2}$Scottish Universities Physics Alliance, Institute for Astronomy, University of Edinburgh, Edinburgh, EH9 3HJ, UK\\
$^{3}$Astrophysics Research Institute, Liverpool John Moores University, Liverpool, L3 5RF, UK\\
$^{4}$Canadian Institute for Advanced Research, 180 Dundas St W, Toronto, ON M5G 1Z8, Canada\\
$^{5}$ Argelander-Institut f\"ur Astronomie, Auf dem H\"ugel 71, D-53121 Bonn, Germany  \\
$^{6}$Canada Research Chair in Observational Cosmology\\
$^{7}$Astrophysics and Cosmology Research Unit,School of Chemistry and Physics, University of KwaZulu-Natal, Durban 4041, South Africa\\
$^{8}$Department of Physics, University of Oxford, Keble Road, Oxford OX1 3RH, UK\\
$^{9}$Leiden Observatory, Leiden University, PO Box 9513, NL-2300 RA Leiden, Netherlands\\
}

\begin{document}

\date{\today}

\pagerange{\pageref{firstpage}--\pageref{lastpage}} \pubyear{2016}

\maketitle

\label{firstpage}

\begin{abstract}
We present measurements of the spatial mapping between (hot) baryons and the total matter in the Universe, via the cross-correlation between the thermal Sunyaev-Zeldovich (tSZ) map from \textit{Planck} and the weak gravitational lensing maps from the Red Sequence Cluster Survey (RCSLenS).  The cross-correlations are performed on the map level where all the sources (including diffuse intergalactic gas) contribute to the signal. We consider two configuration-space correlation function estimators, $\xi^{ y-\kappa}$ and $\xi^ {y-\gamma_{t}}$, and a Fourier space estimator, $C_{\ell}^{y-\kappa}$, in our analysis.  We detect a significant correlation out to three degrees of angular separation on the sky. Based on statistical noise only, we can report 13$\sigma$ and 17$\sigma$ detections of the cross-correlation using the configuration-space $y-\kappa$ and $y-\gamma_{t}$ estimators, respectively. Including a heuristic estimate of the sampling variance yields a detection significance of 6$\sigma$ and 8$\sigma$, respectively. A similar level of detection is obtained from the Fourier-space estimator, $C_{\ell}^{y-\kappa}$.  As each estimator probes different dynamical ranges, their combination improves the significance of the detection. %, as does inclusion of previous CFHTLenS data.  
We compare our measurements with predictions from the cosmo-OWLS suite of cosmological hydrodynamical simulations, where different galactic feedback models are implemented. We find that a model with considerable AGN feedback that removes large quantities of hot gas from galaxy groups and \textit{WMAP}-7yr best-fit cosmological parameters provides the best match to the measurements. All baryonic models in the context of a \textit{Planck} cosmology over-predict the observed signal.  Similar cosmological conclusions are drawn when we employ a halo model with the observed `universal' pressure profile.
\end{abstract} 


%%%%%%%%%%%%%%%%%%%%%%%%%%%%%%%%%%%%%%%%%%%%%%%%%%

%%%%%%%%%%%%%%%%% BODY OF PAPER %%%%%%%%%%%%%%%%%%

\section{Introduction}

Weak gravitational lensing has matured into a precision tool. The fact that it is insensitive to galaxy bias has made lensing a powerful probe of large-scale structure.  However, our lack of a complete understanding of small-scale astrophysical processes has been identified as a major source of uncertainty for the interpretation of the lensing signal. For example, baryonic physics has a significant impact on the matter power spectrum at intermediate and small scales with $k \gtrsim 1 h/\rm{Mpc}$ \citep{vanDaalen2011} and ignoring such effects can lead to significant biases in our cosmological inference \citep{Semboloni2011,Harnois-Deraps2015}. On the other hand, if modelled accurately, these effects can be used as a powerful way to probe the role of baryons in structure formation without affecting the ability of lensing to probe cosmological parameters and the dark matter distribution. 

One can gain insights into the effects of baryons on the total mass distribution by studying the cross-correlation of weak lensing with baryonic probes. In this way, one can acquire information that is otherwise inaccessible, or very difficult to obtain, from the lensing or baryon probes individually.  Cross-correlation measurements also have the advantage that they are immune to residual systematics that do not correlate with the respective signals. This enables the clean extraction of information from different probes.

Recent detections of the cross-correlation between the tSZ signal and gravitational lensing has already revealed interesting insights about the evolution of the density and temperature of baryons around galaxies and clusters. \citet{CFHT1} found a $6~\sigma$ detection of the cross-correlation between the galaxy lensing convergence, $\kappa$, from the Canada-France-Hawaii Telescope Lensing Survey (CFHTLenS) and the tSZ signal ($y$) from \textit{Planck}.  Further theoretical investigations using the halo model \citep{CFHT2} and hydrodynamical simulations (\citealt{CFHT3,Battaglia2015}) demonstrated that $\sim 20\%$ of the cross-correlation signal arises from low-mass halos $M_{\rm{halo}}  \le 10^{14} \msun$, and that about a third of the signal originates from the diffuse gas beyond the virial radius of halos. While the majority of the signal comes from a small fraction of baryons within halos, about half of all baryons reside outside halos and are too cool ($T \sim 10^5K$) to contribute to the measured signal significantly. We also note that \citet{Hill2014} presented a correlation between weak lensing of the CMB (as opposed to background galaxies) and the tSZ with a similar significance of detection, whose signal is dominated by higher-redshift ($z>2$) sources than the galaxy lensing-tSZ signal.

The galaxy lensing-tSZ cross-correlation studies described above were limited. In \citet{CFHT1}, for example, statistical uncertainty dominates due to the relatively small area of the CFHTLenS survey ($\sim 150~{\rm deg}^2$). The tSZ maps were constructed from the first release of the \textit{Planck} data. And finally, the theoretical modeling of the cross-correlation signal was not as reliable for comparison with data as it is today.

In this paper, we use the Red Cluster Survey Lensing (RCSLenS) data \citep{RCSLenS} and the recently released tSZ maps by the \textit{Planck} team \citep{PlancktSZ}. RCSLenS covers an effective area of approximately $560 ~{\rm deg}^2$, which is roughly four times the area covered by CFHTLenS (although the RCSLenS data is somewhat shallower). Combined with the high-quality tSZ maps from \textit{Planck}, we demonstrate a significant improvement in our measurement uncertainties compared to the previous measurements in \citet{CFHT1}. In this paper, we also utilize an estimator of lensing mass-tSZ correlations where the tangential shear is used in place of the convergence. As discussed in Section \ref{sec:formalism}, this estimator avoids introducing potential systematic errors to the measurements during the mass map making process and we also show that it leads to an improvement in the detection significance.

We compare our measurements to the predictions from the cosmo-OWLS suite of cosmological hydrodynamical simulations for a wide range of baryon feedback models. We show that models with considerable AGN feedback reproduce our measurements best when a \textit{WMAP}-7yr cosmology is employed.  Interestingly, we find that all of the models over-predict the observed signal when a \textit{Planck} cosmology is adopted.  In addition, we also compare our measurements to predictions from the halo model with the baryonic gas pressure modelled using the so-called `universal pressure profile' (UPP).  We find consistency in the cosmological conclusions drawn from the halo model approach with that deduced from comparisons to the hydrodynamical simulations.

The organization of the paper is as follows. We present the theoretical background and the data in Section \ref{sec:background}. The measurements are presented in Section \ref{sec:measurement} and the covariance matrix reconstruction is described in Section \ref{sec:covariance}. The implication of our measurements for cosmology and baryonic physics are described in Section \ref{sec:implications} and  we summarize in Section \ref{sec:summary}.

\section{Observational Data and Theoretical Models}
\label{sec:background}

\subsection{Cross-correlation}
\subsubsection{Formalism}
\label{sec:formalism}

We work with two lensing quantities in this paper, the gravitational lensing convergence, $\kappa$, and the tangential shear, $\gamma_t$. The convergence, $\kappa(\thetavc)$ is given by
\be
\kappa(\thetavc) = \int_0^{w_{\rm H}} {\rm d}w \, W^\kappa(w) \, \delta_{\rm m}(\thetavc f_K(w),w),
\label{eq:kappamapdef}
\ee
where $\thetavc$ is the position on the sky,  $w(z)$ is the comoving radial distance to redshift $z$,  $w_{\rm H}$ is the distance to the horizon, and $W^\kappa(w)$ is the lensing kernel \citep{CFHT1},
\be
W^\kappa(w) = {3 \over 2} \Omega_{\rm m} \left({H_0 \over c}\right)^2 g(w) \, {f_K(w) \over a},
\ee
with $\delta_{\rm m}(\thetavc f_K(w),w)$ representing the 3D mass density contrast, $f_K(w)$ is the angular diameter distance at comoving distance $w$, 
and the function $g(w)$ depends on the source redshift distribution $n(w)$ as
\be
g(w) = \int_w^{w_{\rm H}} {\rm d}w' \, n(w') \, {f_K(w'-w) \over f_K(w')}.
\ee

The tSZ signal is due to the inverse Compton scattering of CMB photons off hot electrons  along the line-of-sight which results in a frequency-dependent variation in the CMB temperature \citep{Sunyaev1970},
\be
{\Delta T\over T_0} = y \, S_{\rm SZ}(x),
\ee
where $S_{\rm SZ}(x) = x\coth (x/2) - 4$ is the tSZ spectral dependence, given in terms of $x=h\nu/k_{\rm B} T_0$, $h$ is the Planck constant, $k_{\rm B}$ is the Boltzmann constant, and $T_0=2.725$ K is the CMB temperature. The quantity of interest in the calculations here is the Comptonization parameter, $y$,  given by the line-of-sight integral of the electron pressure:
\be
y(\thetavc) = \int_0^{w_{\rm H}} a \, {\rm d}w \, {k_{\rm B} \sigma_{\rm T} \over m_{\rm e} c^2} \, n_{\rm e} T_{\rm e},
\label{y}
\ee
where $\sigma_{\rm T}$ is the Thomson cross-section, $k_{\rm B}$ is the Boltzmann constant, and $n_{\rm e}[\thetavc f_K(w),w]$ and $T_{\rm e}[\thetavc f_K(w),w]$ are the 3D electron number density and temperature, respectively. 

The first estimator of the tSZ-lensing cross-correlation that we use for the analysis in this paper is the configuration-space two-point cross-correlation function,  $\xi^{y-\kappa}(\vartheta)$:
\be
\xi^{y-\kappa}(\vartheta) = \sum_{\ell} \left(\frac{2\ell + 1}{4\pi}\right ) C_{\ell}^{y-\kappa} \, P_{\ell}(\cos(\vartheta)) \, b^y_{\ell} \, b^{\kappa}_{\ell},
\label{eq:ykappaxi}
\ee
where $P_{\ell}$ are the Legendre polynomials. Note that $\vartheta$ represents the angular separation and should not be confused with the sky coordinate $\thetavc$. The $y-\kappa$ angular cross-power spectrum is 
\be
C_{\ell}^{y-\kappa} = {1 \over 2 \ell +1} \sum_m y_{\ell m} \kappa_{\ell m}^\ast,
\label{eq:ykappaCel}
\ee
where $y_{\ell m}$ and $\kappa_{\ell m}$ are the spherical harmonic transforms of the $y$ and $\kappa$ maps, respectively (see \citealt{CFHT2} for details), and $b^y_{\ell}$ and $b^{\kappa}_{\ell}$ are the smoothing kernels of the $y$ and $\kappa$ maps, respectively.  Note that we ignore higher order lensing corrections to our cross-correlation estimator.  It was shown in \citet{Troster2014} that corrections due to the Born approximation, lens-lens coupling, and higher-order reduced shear estimations have a negligible contribution to our measurement signal.  We also ignore relativistic corrections to the tSZ signal.

Another estimator of lensing-tSZ correlations is constructed using the tangential shear, $\gamma_t$, which is defined as 
\be
%<\gamma_t ( \theta)> = \bar{\kappa} (< \theta) - <\kappa> (\theta)
\gamma_t (\thetavc) = - \gamma_1 \cos(2\phi) - \gamma_2 \sin(2\phi) ,
\ee
where ($\gamma_1$,$\gamma_2$) are the shear components relative to Cartesian coordinates, $\thetavc = [\vartheta~\cos(\phi),\vartheta~\sin(\phi)]$ where $\phi$ is the polar angle of $\thetavc$ with respect to the coordinate system. In the flat sky approximation, the Fourier transform of $\gamma_t$ can be written in terms of the Fourier transform of the convergence as \citep{Jeong2009}:

\begin{equation}
 \gamma_t({\thetavc}) = -\int
  \frac{d^2\mathbf{l}}{(2\pi)^2}\kappa(\mathbf{l})\cos[2(\phi-\varphi)]e^{il\theta\cos(\phi-\varphi)}.
\end{equation}
where $\varphi$ is the angle between $\mathbf{l}$ and the cartesian coordinate system. We use the above expression to derive the $y-\gamma_t$ cross-correlation function as

\begin{equation}
\label{eq:xigammaty}
\xi^{y-\gamma_t}(\vartheta) = <y~\gamma_t> (\vartheta)=  \int
  \frac{d^2 \mathbf{l}}{(2\pi)^2} C_{\ell}^{y\kappa}  \cos[2(\phi-\varphi)]e^{il\vartheta\cos(\phi-\varphi)} .
  \end{equation}
Note that the correlation function that we have introduced in Eq.~\ref{eq:xigammaty} differs from what is commonly used in galaxy-galaxy lensing studies, where the average shear profile of halos $<{\gamma}_t>$ is measured:
\begin{equation}
<{\gamma}_t> (\vartheta) \equiv \int_0^{2\pi}\frac{d\phi}{2\pi}
\gamma_t(\vartheta,\phi).
\end{equation}
Here, we take every point in the $y$ map, compute the corresponding tangential shear from every galaxy at angular separation $\vartheta$ in the shear catalog and then take the average (instead of computing the signal around identified halos).  Working with the shear directly in this way, instead of convergence, has the advantage that we skip the mass map reconstruction process and any noise and systematic issues that might be introduced during the process.  We have successfully applied similar estimators previously to compute the cross-correlation of galaxy lensing with CMB lensing in \citet{Harnois-Deraps2016}. In principle, this estimator can be used for cross-correlations with any other scalar quantity.
 
\subsubsection{Fourier-space versus configuration-space analysis}
\label{sec:Fourier}

In addition to the configuration-space analysis described above, we also study the cross-correlation in the Fourier space.  A configuration-space analysis has the advantage that there are no complications introduced by the presence of masks, which significantly simplifies the analysis. As described in \citet{Harnois-Deraps2016}, a Fourier analysis requires extra considerations to account for the impact of several factors, including the convolution of the mask power spectrum and mode-mixing.  On the other hand, a Fourier space analysis can be useful in distinguishing between different physical effects at different scales (e.g., the impact of baryon physics and AGN feedback).  We choose a forward modeling approach as described in \citet{Harnois-Deraps2016} and discussed further in Section \ref{sec:measurement}.

\subsection{Observational data}
\label{sec:data}

\subsubsection{RCSLenS lensing maps}

The Red Sequence Cluster Lensing Survey \citep{RCSLenS} is part of the second Red-sequence Cluster Survey (RCS2; \citet{Gilbank:2010zv})\footnote{The RCSLenS data are public and can be found at: {\tt www.rcslens.org}}. Data was acquired from the MegaCAM camera from 14 separate fields and covers a total area of 785 deg$^2$ on the sky. %A similar pipeline  was used to process both RCSLenS and CFHTLenS data. 
The pipeline used to process RCSLenS data includes a reduction algorithm \citep{Erben2013}, followed by photometric redshift estimation \citep{Hildebrandt2012,Benitez2000} and a shape measurement algorithm \citep{Miller2013}. For a complete description see \citet{Heymans2012} and \citet{RCSLenS}.

For some of the RCSLenS fields the photometric information is incomplete, so we use external data to estimate the galaxy source redshift distribution, $n(z)$.  The CFHTLenS-VIPERS photometric sample is used which contains near-UV and near-IR data combined with the CFHTLenS photometric sample and is calibrated against $\sim$ 60000 spectroscopic redshifts \citep{Coupon2015}. The source redshift distribution, $n(z)$, is then obtained by stacking the posterior distribution function of the CFHTLenS-VIPERS galaxies with predefined magnitude cuts and applying the following fitting function (following the procedure outlined in Section 3.1.2 of \citet{Harnois-Deraps2016}):
\begin{eqnarray}
\label{eq:nofz}
n_{\rm{RCSLenS}}(z) &=&  a~z ~\exp {[\frac{-(z-b)^2}{c^2}]} +  d~z  ~\exp {[\frac{-(z-e)^2}{f^2}]} \nonumber \\
&+& g~z  ~\exp {[\frac{-(z-h)^2}{i^2}]} .  
\end{eqnarray}
As described in the Appendix \ref{sec:extras}, we experimented with several different magnitude cuts to find the range where the SNR for our measurements is maximized.  We find that selecting galaxies with $\rm{mag}_r > 18$ yields the highest SNR with the best-fit values of ($a, b, c, d, e, f, g, h, i$) = (2.94, -0.44, 1.03, 1.58, 0.40, 0.25, 0.38, 0.81, 0.12). This cut leaves us with approximately 10 million galaxies from the 14 RCSLenS fields, yielding an effective surface number density of $\bar{n} = 5.8$ gal/arcmin$^{2}$ and an ellipticity dispersion of $\sigma_{\epsilon} = 0.277$.

\begin{figure}
{\centering{
\includegraphics[width=1.0\columnwidth]{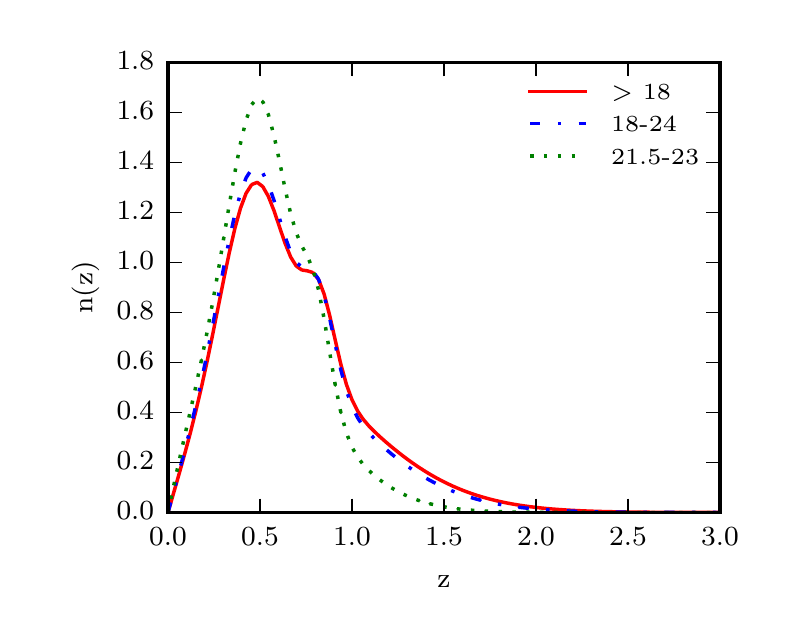}
}}
\caption{Redshift distribution, $n(z)$, of the RCSLenS sources for different r-magnitude cuts. We work with the $\rm{mag}_r > 18$ cut  (which includes all the objects in the survey).% The $n(z)$ for the CFHTLenS is also over-plotted for comparison. 
}
\label{fig:nofz}
\end{figure}

Fig.~\ref{fig:nofz} shows the source redshift distributions $n(z)$ for the three different magnitude cuts we have examined. Note that the lensing signal is most sensitive in the redshift range approximately half way between the sources and the observer. RCSLenS is shallower than the CFHTLenS (see the analysis in \citet{vanWaerbeke2013}) but, as we demonstrate later, the larger area coverage of RCSLenS (more than) compensates for the lower number density of the source galaxies, in terms of the measurement of the cross-correlation with the tSZ signal. 

%**************************************
%**************************************

For our analysis we use the shear data as well as the reconstructed projected mass maps (convergence maps) from RCSLenS. For the tSZ-tangential shear cross-correlation ($y-\gamma_t$), we work at the catalogue level where each pixel in the $y$ map is correlated with the average tangential shear from the corresponding shear data in an annular bin around that point, as described in Section \ref{sec:configuration-space}. To construct the convergence maps, we follow the method described in \citet{vanWaerbeke2013}. In Appendix \ref{sec:extras} we study the impact of map smoothing on the signal to noise ratio (SNR) we determine for the $y-\kappa$ cross-correlation analysis. We demonstrate that the best SNR is obtained when the maps are smoothed with a kernel that roughly matches the beam scale of the corresponding $y$ maps from \textit{Planck} survey (FWHM = 9.5 arcmin). 

The noise properties of the constructed maps are studied in detail in Appendix \ref{sec:null}. 

\subsubsection{\textit{Planck} tSZ y maps}

For the cross-correlation with the tSZ signal, we use the full sky maps provided in the \textit{Planck} 2015 public data release \citep{PlancktSZ}.  We use the \emph{milca} map that has been constructed from multiple frequency channels of the survey.  Since we are using the public data from the {\it Planck} collaboration, there is no significant processing involved.  Our map preparation procedure is limited to masking the map and cutting the patches matching the RCSLenS footprint. 

Note that in performing the cross-correlations we are limited by the footprint area of the lensing surveys.  In the case of RCSLenS, we have 14 separate compact patches with different sizes.  In contrast, the tSZ $y$ maps are full-sky (except for masked regions).  We therefore have the flexibility to cut out larger regions around the RCSLenS fields, in order to provide a larger cross-correlation area that helps suppress the statistical noise, leading to an improvement in the SNR.  We cut out $y$ maps so that there is complete overlap with RCSLenS up to the largest angular separation in our cross-correlation measurements.  

Templates have also been released by the {\it Planck} collaboration to remove various contaminating sources. We use their templates to mask galactic emission and point sources, which amounts to removing $\sim$40\% of the sky. We have compared our cross-correlation measurements with and without the templates and checked that our signal is robust.  We have also separately checked that the masking of point sources has a negligible impact on our cross-correlation signal (see Appendix \ref{sec:extras}).  These sources of contamination do not bias our cross-correlation signal and contribute only to the noise level.

In addition to using the tSZ map from the \textit{Planck} collaboration, we have also tested our cross-correlation results with the maps made independently following the procedure described in \citet{CFHT1}, where several full-sky $y$ maps were constructed from the second release of \textit{Planck} CMB band maps. To construct the maps, a linear combination of the four HFI frequency band maps (100, 143, 217, and 353 GHz) were used and smoothed with a Gaussian beam profile with $\theta_{\rm SZ, FWHM} = 9.5$ arcmin.  To combine the band maps, band coefficients were chosen such that the primary CMB signal is removed, and the dust emission with a spectral index $\beta_{\rm d}$ is nulled.  A range of models with different  $\beta_{\rm d}$ values were employed to construct a set of $y$ maps that were used as diagnostics of residual contamination.  The resulting cross-correlation measurements vary by roughly 10\% between the different $y$ maps, but are consistent within the errors with the measurements from the public \textit{Planck} map.  

\subsection{Theoretical models}
\label{sec:models}

We compare our measurements with theoretical predictions based on the halo model and from full cosmological hydrodynamical simulations.  Below we describe the important aspects of these models.

\subsubsection{Halo model}
\label{sec:halo-model}

We use the halo model description for the tSZ - lensing cross-correlation developed in \citet{CFHT2}.  In the framework of the halo model, the $y-\kappa$ cross-correlation power spectrum is: 
\begin{equation}
C^{y-\kappa}_{\ell} = C_{\ell}^{y-\kappa,\textrm{1h}}
 + C_{\ell}^{y-\kappa,\textrm{2h}} \, ,
\label{eq:cell-kappay}
\end{equation}
where the 1-halo and 2-halo terms are defined as 
\begin{eqnarray}
C_{\ell}^{y-\kappa,\textrm{1h}}  &=&  \int^{z_{\rm max}}_{0}{\rm d}z
 \frac{{\rm d}V}{{\rm d}z{\rm d}\Omega}   \int^{M_{\rm max}}_{M_{\rm min}}{\rm d}M
 \frac{\der n}{\der M} y_{\ell}(M,z)\, \kappa_{\ell}(M,z),  \nonumber \,  \\
 C^{y-\kappa,\textrm{2h}}_{\ell}  &= &  \int^{z_{\rm max}}_{0} \,
 \der z \frac{\der V}{\der z \der \Omega}\,P^{\rm lin}_{\rm m}(k=\ell/\chi,z) \nonumber \\ 
& \times & \left[\int^{M_{\rm max}}_{M_{\rm min}}\der M
 \frac{\der n}{\der M}b(M,z) \kappa_{\ell}(M,z) \right] \nonumber  \\
& \times & \left[\int^{M_{\rm max}}_{M_{\rm min}}\der M
  \frac{\der n}{\der M}b(M,z) y_{\ell}(M,z) \right] . 
  \label{eq:halo}
\end{eqnarray}
In the above equations $P^{\rm lin}_{\rm m}(k,z)$ is the 3D linear matter power
spectrum at redshift $z$, $\kappa_{\ell}(M,z)$ is the Fourier transform of the convergence
profile of a single halo of mass $M$ at redshift $z$ with the NFW profile:
\begin{equation}
\kappa_{\ell}=\frac{W^{\kappa}(z)}{\chi^{2}(z)}
 \frac{1}{\bar{\rho}_{\rm m}}\int^{r_{\rm vir}}_{0}\der r (4 \pi r^{2})
 \frac{\sin(\ell r/\chi)}{\ell r/\chi} \rho(r;M,z), \label{eq:kappa-ell}
\end{equation}
and $y_{\ell}(M,z)$ is the Fourier
transform of the projected gas pressure profile of a single halo:
\begin{equation}
 y_{\ell}=\frac{4\pi r_{\rm s}}{\ell^{2}_{\rm s}}
 \frac{\sigma_{\rm T}}{m_{\rm e}c^{2}} \int^{\infty}_{0} \der x \, x^{2}
 \frac{\sin(\ell x/\ell_{s})}{\ell x/\ell_{s}}
 P_{\rm e}(x;M,z). \label{eq:y-ell}
\end{equation}
Here $x \equiv a(z)r/r_{\rm s}$ and $\ell_{\rm s}=a\chi/r_{\rm s}$, where
$r_{\rm s}$ is the scale radius of the 3D pressure profile, and $P_{\rm e}$ is the 3D electron pressure. The ratio $r_{\rm vir}/r_{\rm s}$ is the concentration parameter (see e.g \citet{CFHT2} for details). 

%%%%%%%%%%%%%%%%%%%%%%%%%%%
\begin{table*}
\centering
\caption{Sub-grid physics of the baryon feedback models in the cosmo-OWLS runs.  Each model has been run adopting both the {\it WMAP}-7 and {\it Planck} cosmologies.}
\begin{tabular}{|l|l|l|l|l|l|l|}
         \hline
	Simulation & UV/X-ray background & Cooling & Star formation & SN feedback & AGN feedback & $\Delta T_{\rm heat}$ \\
	\hline
        NOCOOL & Yes & No & No & No & No & ...\\
        REF & Yes & Yes & Yes & Yes & No & ...\\
        AGN 8.0 & Yes & Yes & Yes & Yes & Yes & $10^{8.0}$ K\\
        AGN 8.5 & Yes & Yes & Yes & Yes & Yes & $10^{8.5}$ K\\
        AGN 8.7 & Yes & Yes & Yes & Yes & Yes & $10^{8.7}$ K\\
        \hline
\end{tabular}
\label{table:cosmo_owls}
\end{table*} 

%%%%%%%%%%%%%%%%%%%%%%%%%%%

For the electron pressure of the gas in halos, we adopt the so-called `universal pressure profile' (UPP; \citealt{Arnaud2010}):
\begin{eqnarray}
P(x\equiv r/R_{500}) &=& 1.65 \times 10^{-3} E(z)^{\frac{8}{3}}
 \left(\frac{M_{500}}{3\times 10^{14}h_{70}^{-1}\msun}
 \right)^{\frac{2}{3}+0.12} \nonumber \\
 &\times& \mathbb{P}(x) \, h_{70}^{2}\text{ }\left[\textrm{keV cm}^{-3}\right],
 \label{eq:upp}
\end{eqnarray}
where $\mathbb{P}(x)$ is the generalized NFW model \citep{Nagai2007}:
\begin{equation}
\mathbb{P}(x) = \frac{P_0}{(c_{500} x)^{\gamma}
 \left[1+(c_{500} x)^{\alpha}\right]^{(\beta-\gamma)/\alpha}}. \label{px}
\end{equation}
We use the best-fit parameter values from \citet{Planck2013}: $\{P_{0}, c_{500}, \alpha, \beta, \gamma\} = \{6.41, 1.81, 1.33, 4.13, 0.31\}$. To compute the configuration-space correlation functions, we use Eqs.~\ref{eq:ykappaxi} and \ref{eq:xigammaty} for $\xi^{y-\kappa}$ and  $\xi^{y-\gamma_t}$, respectively. We present the halo model predictions for two sets of cosmological parameters: the maximum-likelihood \textit{Planck} 2013 cosmology \citep{Planck2014} and the maximum-likelihood \textit{WMAP}-7yr cosmology \citep{Komatsu2011} with \{$\Omega_{\rm m}$, $\Omega_{\rm b}$, $\Omega_{\Lambda}$, $\sigma_8$, $n_{\rm s}$, $h$\} = {0.3175, 0.0490, 0.6825, 0.834, 0.9624, 0.6711} and \{0.272, 0.0455, 0.728, 0.81, 0.967, 0.704\}, respectively. 

There are several factors that have an impact on these predictions; the choice of the gas pressure profile, the adopted cosmological parameters, and the $n(z)$ distribution of sources in the lensing survey. In addition, the hydrostatic mass bias parameter, $b$ (defined as $M_{\rm{obs,500}}=(1-b) M_{\rm{true,500}}$), alters the relation between the adopted  pressure profile and the true halo mass.  Typically, it has been suggested that $1-b \approx 0.8$. %  We assume this bias by default in our halo model calculations and note that any deviation from this value 
Note that the impact of the hydrostatic mass bias in real groups and clusters will be absorbed into our amplitude fitting parameter $A_{\rm{tSZ}}$ (defined in Eq.~\ref{eq:Atsz}).

\subsection{The cosmo-OWLS hydrodynamical simulations}
\label{sec:simulations}

We also compare our measurements to predictions from the cosmo-OWLS suite of hydrodynamical simulations. In \citet{CFHT3} we compared these simulations to measurements using CFHTLenS data and we also demonstrated that high resolution tSZ-lensing cross-correlations have the potential to simultaneously constrain cosmological parameters and baryon physics.  Here we build on our previous work and employ the cosmo-OWLS simulations in the modelling of RCSLenS data.

The cosmo-OWLS suite is an extension of the OverWhelmingly Large Simulations project (OWLS; \citealt{Schaye2010}). The suite consists of box-periodic hydrodynamical simulations with volumes of $(400 \ h^{-1} \ {\rm Mpc})^3$ and $1024^3$ baryon and dark matter particles.  The initial conditions are based on either the {\it WMAP}-7yr or {\it Planck}~2013 cosmologies. We quantify the agreement of our measurements with the predictions from each cosmology in Section \ref{sec:implications} . 

We use five different baryon models from the suite as summarized in Table \ref{table:cosmo_owls} and described in detail in \citet{LeBrun2014} and \citet{McCarthy2014} and references therein. \textsc{NOCOOL} is a standard non-radiative (`adiabatic') model. \textsc{REF} is the OWLS reference model and includes sub-grid prescriptions for star formation \citep{Schaye2008}, metal-dependent radiative cooling \citep{Wiersma2009a}, stellar evolution, mass loss, chemical enrichment \citep{Wiersma2009b}, and a kinetic supernova feedback prescription \citep{DallaVecchia2008}. The \textsc{AGN} models are built on the \textsc{REF} model and additionally include a prescription for black hole growth and feedback from active galactic nuclei \citep{Booth2009}. The three \textsc{AGN} models differ only in their choice of the key parameter of the AGN feedback model $\Delta T_{\rm heat}$, which is the temperature by which neighbouring gas is raised due to feedback.  Increasing the value of $\Delta T_{\rm heat}$ results in more energetic feedback events, and also leads to more bursty feedback, since the black holes must accrete more matter in order to heat neighbouring gas to a higher adiabat.

Following \citet{McCarthy2014}, we produce light cones of the simulations by stacking randomly rotated and translated simulation snapshots (redshift slices) along the line-of-sight back to $z=3$. Note that we use 15 snapshots at fixed redshift intervals between $z=0$ and $z=3$ in the construction of the light cones. This ensures a good comoving distance resolution, which is required to capture the evolution of the halo mass function and the tSZ signal.  The light cones are used to produce $5\times 5$ degree maps of the $y$, shear ($\gamma1$, $\gamma2$) and convergence ($\kappa$) fields. We construct 10 different light cone realizations for each feedback model and for the two background cosmologies.  Note that in the production of the lensing maps we adopt the source redshift distribution, $n(z)$, from the RCSLenS survey to produce a consistent comparison with the observations.  

From our previous comparisons to the cross-correlation of CFHTLenS weak lensing data with the initial public \textit{Planck} data in \citet{CFHT3}, we found that the data mildly preferred a \textit{WMAP}-7yr cosmology to the \textit{Planck}~2013 cosmology.  We will revisit this in Section \ref{sec:implications} in the context of the new RCSLenS data.

\section{Observed cross-correlation}
\label{sec:measurement}

Below we describe our cross-correlation measurements between tSZ $y$ and galaxy lensing quantities using the configuration-space and Fourier-space estimators described in Section \ref{sec:formalism}. 

\subsection{Configuration-space analysis}
\label{sec:configuration-space}

%**************************************
\begin{figure*}
{\centering{
\includegraphics[width=1.0\columnwidth]{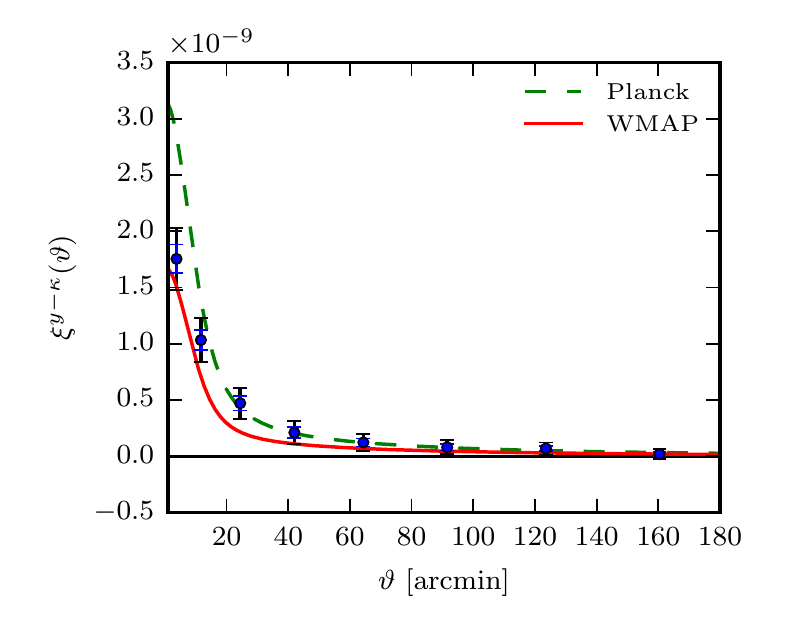}
\includegraphics[width=1.0\columnwidth]{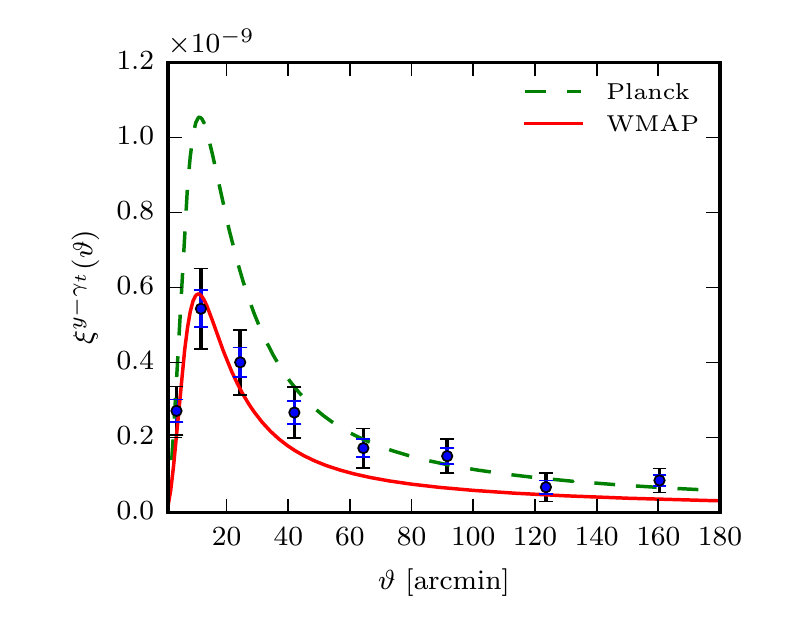}
}}
\caption{Cross-correlation measurements of $y-\kappa$ (left) and $y-\gamma_t$ (right) from RCSLenS. The larger (smaller) error bars represent uncertainties after (before) including our estimate of the sampling variance contribution (see Section \ref{sec:covariance}). Halo model predictions using UPP with \textit{WMAP}-7yr and \textit{Planck} cosmologies are also over-plotted for comparison. 
}
\label{fig:RCS-A}
\end{figure*}
%**************************************

We perform the cross-correlations on the 14 RCSLenS fields. The measurements from the fields converge around the mean values at each bin of angular separation with a scatter that is mainly dominated by sampling variance (see Section \ref{sec:error-inflate}).  To combine the fields, we take the weighted mean of the field measurements, where the weights are determined by the unmasked area of the fields. 

As described earlier, to improve the SNR and suppress statistical noise, we use ``extended" $y$ maps around each RCSLenS field to increase the cross-correlation area. For RCSLenS, we extend our measurements to an angular separation of 3 degrees, and hence include 4 degree wide bands around the RCSLenS fields.  

We compute our configuration-space estimators as described below. For $y-\gamma_t$, we work at the catalogue level and compute the two-point correlation function as:

\begin{equation}
	\label{eq:gammat-estimator}
	\xi^{y-\gamma_t}(\vartheta) = \frac{\sum_{ij} y^i e_t^{ij} w^j \Delta_{ij}(\vartheta)}{\sum_{ij}  w^j \Delta_{ij}(\vartheta)}\frac{1}{1+K(\vartheta)} \,,
\end{equation}
where $y^i$ is the value of pixel $i$ of the tSZ map, $e_t^{ij}$ is the tangential ellipticity of galaxy $j$ in the catalog with respect to pixel $i$, and $w^j$ is the {\it lens}fit weight (see \citealt{Miller2013} for technical definitions). 
The $(1+K(\vartheta))^{-1}$ factor accounts for the multiplicative calibration correction (see \citet{RCSLenS} for details): %  \citep{Heymans2012c}:
 \begin{equation}
	\frac{1}{1+K(\vartheta)} = \frac{\sum_{ij}  w^j \Delta_{ij}(\vartheta)}{\sum_{ij}  w^j (1+m^j)\Delta_{ij}(\vartheta)} \ .
\end{equation}
Finally, $\Delta_{ij}(\vartheta)$ is imposes our binning scheme and is $1$ if the angular separation is inside the bin centered at $\vartheta$ and zero otherwise.

For the $y-\kappa$ cross-correlation, we use the corresponding mass maps for each field and compute the correlation function as:

\begin{equation}
\label{eq:kappa-estimator}
	\xi^{y-\kappa}( \vartheta) = \frac{\sum_{ij} y^i \kappa^j \Delta_{ij}(\vartheta)}{\sum_{ij}  \Delta_{ij} (\vartheta)}. 
\end{equation}
where $\kappa^j $ is the convergence value at pixel $j$ and includes the necessary weighting, $w^j$. 

Fig.~\ref{fig:RCS-A} presents our configuration-space measurement of the RCSLenS cross-correlation with \textit{Planck} tSZ. The filled circle data points show the $y-\kappa$ (left) and $y-\gamma_t$ (right) cross-correlations.  To guide the eye, the solid red curves and dashed green curves represent the predictions of the halo model for \textit{WMAP}-7yr and \textit{Planck}~2013 cosmologies, respectively. We have 8 square root-spaced bins between 1 and 180 arcminutes.

\subsection{Fourier-space measurements}
\label{sec:Fourier-space}

\begin{figure}
{\centering{
\includegraphics[width=1.0\columnwidth]{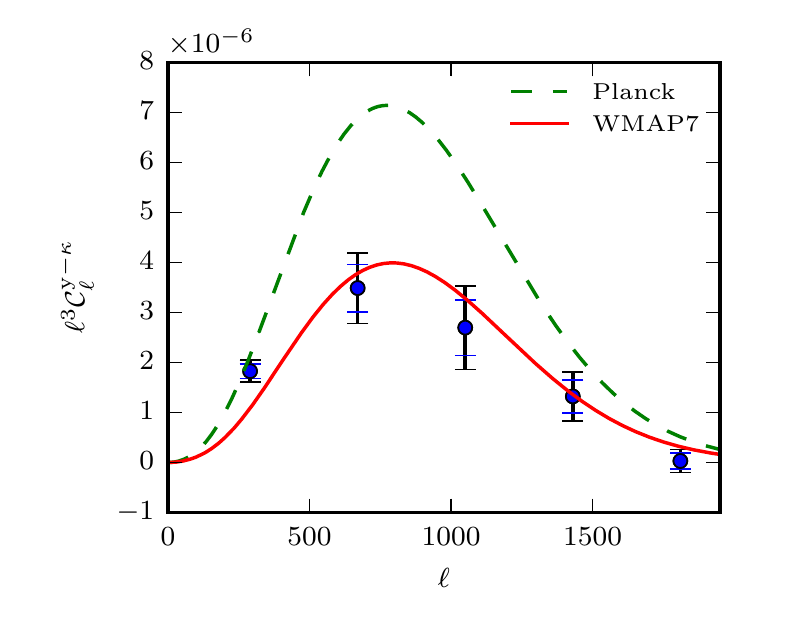}
}}
\caption{Similar to Fig. \ref{fig:RCS-A} but for Fourier-space estimator,  $C_{\ell}^{y-\kappa}$.}
\label{fig:cell_measurement}
\end{figure}

In the Fourier-space analysis, we work with the convergence and tSZ maps.  As detailed in \citet{Harnois-Deraps2016}, it is important to account for a number of numerical and observational effects when performing the Fourier-space analysis. These effects include data binning, map smoothing, masking, zero-padding and apodization. Failing to take such effects into account will bias the cross-correlation measurements significantly.

Here we adopt the forward modeling approach described in \citet{Harnois-Deraps2016}, where theoretical predictions are turned into a `pseudo-$C_{\ell}$', as summarized below.  First, we obtain the theoretical $C_{\ell}$ predictions from Eqs.~(\ref{eq:cell-kappay}) and (\ref{eq:halo}) as described in Section \ref{sec:halo-model}. We then multiply the predictions by a Gaussian smoothing kernel that matches the Gaussian filter used in constructing the $\kappa$ maps in the mass map making process, and another smoothing kernel that accounts for the beam effect of the \textit{Planck} satellite.

Next we include the effects of observational masks on our power spectra which breaks down into three components (see \citet{Harnois-Deraps2016} for details): i) an overall downward shift of power due to the masked pixels which can be corrected for with a rescaling by the number of masked pixels; ii) an optional apodization scheme that we apply to the masks to smooth the sharp features introduced in the power spectrum of the masked map that enhance the high-$\ell$ power spectrum measurements; and iii) a mode mixing matrix, that propagates the effect of mode coupling due to the observational window. 

As shown in \citet{Harnois-Deraps2016}, steps (ii) and (iii) are not always necessary in the context of cross-correlation when the masks from both maps do not strongly correlate with the data. We have checked that this is indeed the case by measuring the cross-correlation signal from the cosmo-OWLS simulations with and without applying different sections of the RCSLenS masks, with and without apodization, and observed that changes in the results were minor.  We therefore choose to remove the steps (ii) and (iii) from the analysis pipeline. As the last step in our forward modeling, we re-bin the modeled pseudo-$C_{\ell}$ so that it matches the binning scheme of the data. Note that these steps have to be calculated separately for each individual field due to their distinct masks.

Fig.~\ref{fig:cell_measurement} shows our Fourier-space measurement for the $y-\kappa$ cross-correlation, where halo model predictions for the \textit{WMAP}-7yr and \textit{Planck} cosmologies are also over-plotted. %We have 5 linearly-spaced bins between $\ell=1$ and $\ell=$2000. 
Our Fourier-space measurement is consistent with the configuration-space measurement overall. Namely, the data points provide a better match to \textit{WMAP}-7yr cosmology prediction on small physical scales (large $\ell$ modes) and tend to move towards the \textit{Planck} prediction on large physical scales (small $\ell$ modes). A more detailed comparison is non-trivial as different scales ($\ell$ modes) are mixed in the configuration-space measurements.  

The details of error estimation and the significance of the detection are described in Section \ref{sec:covariance}.

\section{Estimation of covariance matrices and significance of detection }
\label{sec:covariance}

In this section, we describe the procedure for constructing the covariance matrix and the statistical analysis that we perform to estimate the significance of our measurements. We have investigated several methods for estimating the covariance matrix for the type of cross-correlations performed in this paper.

\subsection{Configuration-space covariance}

To estimate the covariance matrix we follow the method of \citet{vanWaerbeke2013}. We first produce 300 random shear catalogues from each of the RCSLenS fields. We create these catalogs by randomly rotating the individual galaxies. This procedure will destroy the underlying lensing signal and create catalogs with pure statistical lensing noise. We then construct the $y-\gamma_t$ covariance matrix, $\mathcal{C}^{y-\gamma_t}$,  by cross-correlating the randomized shear maps for each field with the $y$ map. 

To construct the $y-\kappa$ covariance matrix, we perform our standard mass reconstruction procedure on each of the 300 random shear catalogs to get a set of convergence noise maps. We then compute the covariance matrix by cross-correlating the $y$ maps with these random convergence maps. We follow the same procedure of map making (masking, smoothing etc) in the measurements from random maps as we did for the actual measurement. This ensures that our error estimation is representative of the underlying covariance matrix. %Finally, we repeat the procedure to produce the covariance matrices for CFHTLenS.  

Note that we also need to ``debias" the inverse covariance matrix by a debiasing factor as described in \citet{Hartlap2007}:
\begin{equation}
\alpha = (n-p-2)/(n-1) \, ,
\label{eq:Hartlap}
\end{equation} 
where $p$ is the number of data bins and $n$ is the number of random maps used in the covariance estimation\footnote{In principle we should also implement the treatment of \citet{Sellentin2016}, but the precision of our measurement is not high enough to worry about such errors.}. 

%**************************************
\begin{figure*}
{\centering{
\includegraphics[width=1.\columnwidth]{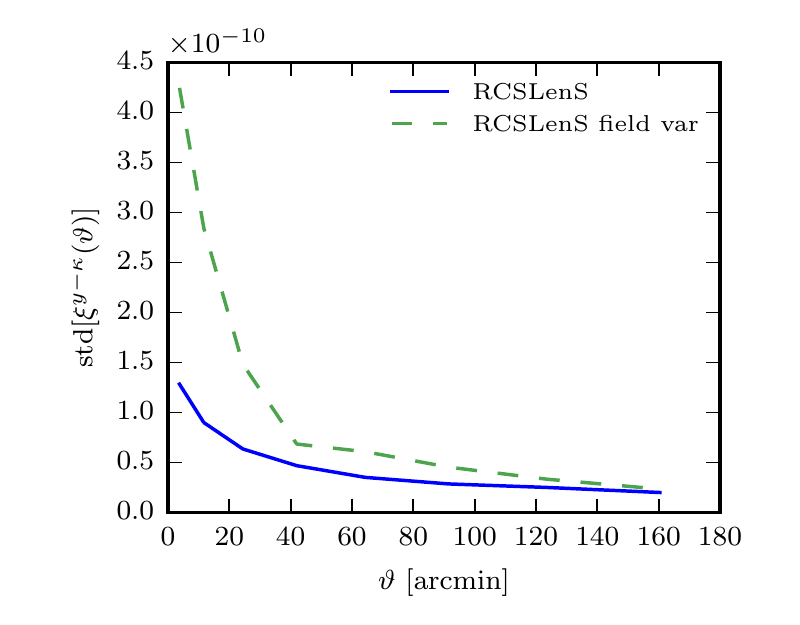}
\includegraphics[width=1.\columnwidth]{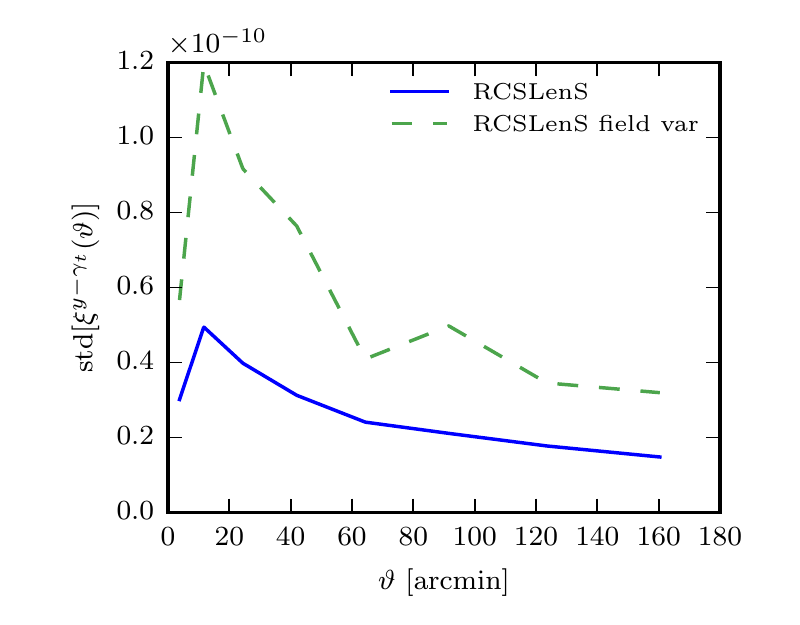}
}}
\caption{Comparison of the errors (from the diagonal elements of the covariance matrix) of the $y-\kappa$ (left) and $y-\gamma_t$ (right) measurements from RCSLenS. 
Over-plotted with dashed lines are the standard deviations estimated from the scatter of the cross-correlation signal from individual fields in the survey that include contribution from statistical noise and sampling variance. We inflate our estimated errors to match the measured standard deviations so that we can take into account the impact of sampling variance (see Section \ref{sec:error-inflate}). }
\label{fig:errors-RCSvsCFHT}
\end{figure*}
%**************************************

\begin{figure*}
{\centering{
\includegraphics[width=0.65\columnwidth]{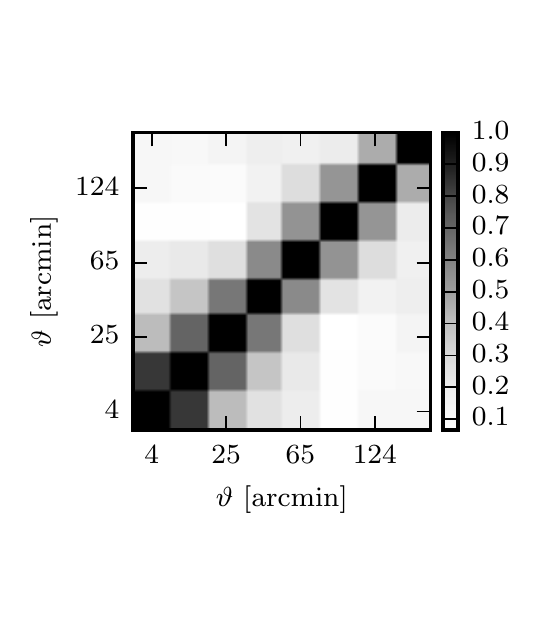}
\includegraphics[width=0.65\columnwidth]{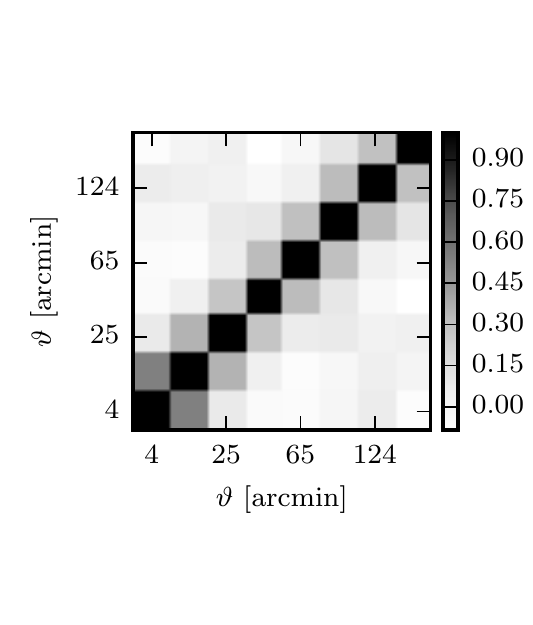}
\includegraphics[width=0.65\columnwidth]{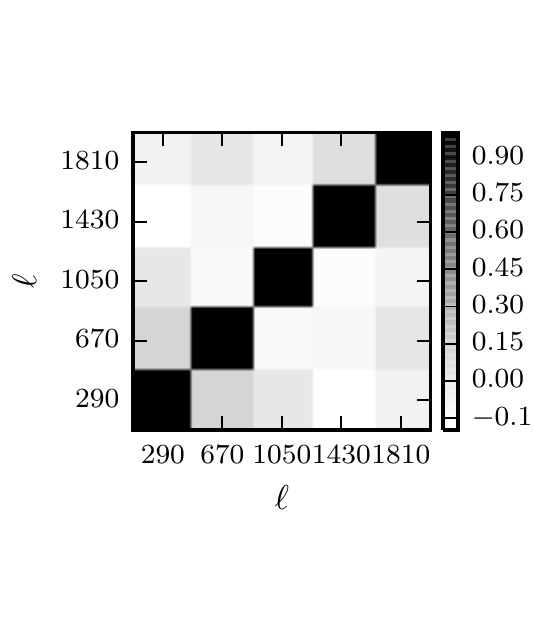}
}}
\caption{Cross-correlation coefficient matrix of the angular bins for the configuration-space ${y-\kappa}$ (left) and $y-\gamma_t$ (middle), and the Fourier-space ${y-\kappa}$ (right) estimators. Angular bins are more correlated for the $y-\kappa$ estimator compared to $y-\gamma_t$ or the Fourier-space estimator. }
\label{fig:correlation}
\end{figure*}

The correlation coefficients are shown in Fig. \ref{fig:correlation} for $y-\kappa$ (left) and $y-\gamma_t$ (right). As a characteristic of configuration-space, there is a high level of correlation between pairs of data points within each estimator. This is more pronounced for $y-\kappa$ since the mass map construction is a non-local operation, and also that the maps are smoothed which creates correlation by definition. Having a lower level of bin-to-bin correlations is another reason why one might want to work with tangential shear measurements rather than mass maps in such cross-correlation studies.

%**************************************
%**************************************

\subsection{Estimating the contribution from the sampling variance}
\label{sec:error-inflate}

Constructing the covariance matrix as described above includes the statistical noise contribution only. There is, however, a considerable scatter in the cross-correlation signal from the individual fields. A comparison of the observed scatter to that among different LoSs of the (noise-free) simulations shows that the majority of this scatter is due to sampling variance. We therefore need to include the contribution to the covariance matrix from sampling variance. 

We are not able, however, to estimate a reliable covariance matrix that includes sampling variance since the number of samples we have access to is very limited. We only have a small number of fields from the lensing surveys (14 fields from RCSLenS is not nearly enough) and the same is true for the number of LoS maps from hydrodynamical simulations (10 LoS). Instead, we can estimate the sampling variance contribution by quantifying by how much we need to ``inflate'' our errors to account for the impact of sampling variance.

The scatter in the cross-correlation signal from the individual fields is due to both statistical noise and sampling variance. We compare the scatter (or variance) in each angular bin to that of the diagonal elements of the reconstructed covariance matrix that we obtained from the previous section (which quantifies the statistical uncertainty alone). Fig. \ref{fig:errors-RCSvsCFHT} overplots the standard deviation of the scatter of the cross-correlation signal from individual fields for the two estimators. We estimate the scaling factor by which we should inflate the diagonal elements of the computed covariance matrix to match the observed scatter. Encouragingly, when rescaled, the variance and diagonal elements have roughly the same shape. We use these best-fit scaling factor to inflate the errors as an estimate of the full error budget. Table \ref{table:error-inflate} summarizes the best-fit scaling factors for both estimators. % and for the two lensing surveys. 

%**************************************
\begin{table}
\centering
\caption{The best-fit scaling factor that inflates the estimated errors to match the observed scatter in the cross-correlation amplitude from individual fields.}
\begin{tabular}{|l|l|l|l|l|l|l|}
 \hline
  Estimator:	 			  & $y-\kappa$ & $y-\gamma_t$ & Combined   \\
 \hline
 \hline					  
%CFHTLenS       &  1.5 & 1.6 & 1.8 \\
%\hline
%RCSLenS        
Config. space: & 2.1 & 2.2 & 2.1 \\
Fourier space: &  1.5 & N/A & N/A \\
\hline
\end{tabular}
\label{table:error-inflate}
\end{table} 
%**************************************

\subsection{Fourier-space covariance}

For the covariance matrix estimation in Fourier space, we follow a similar procedure as in configuration space. We first Fourier transform the random convergence maps, and then follow the same analysis for the measurements (see Section \ref{sec:measurement}). The resulting cross-correlation measurements create a large sample that can be used to construct the covariance matrix.  Similar to the configuration space analysis, we also debias the computed covariance matrix. 

Fig. \ref{fig:correlation}, right shows the cross-correlation coefficients for the $\ell$ bins (Note that we chose to work with 5 linearly-spaced bins between $\ell=1$ and $\ell=2000$). As expected, there is not much bin-to-bin correlation and the off-diagonal elements are small. 

\subsection{$\chi^2$ analysis and significance of detection}
\label{sec:significance}

We quantify the significance of our measurements using the SNR estimator as described below. We assume that the RCSLenS fields are sufficiently separated such that they can be treated as independent, ignoring field-to-field covariance. 

First, we introduce the cross-correlation bias factor, $A_{tSZ}$, through:
\begin{equation}
\mathcal{V} = \tilde{\xi} -  A_{\rm{tSZ}}~ \hat{\xi}.
\label{eq:Atsz}
\end{equation}
$A_{\rm{tSZ}}$ quantifies the difference in amplitude between the measured ($\tilde{\xi}$) and predicted ($\hat{\xi}$) cross-correlation function. The prediction can be from either the halo model or from hydrodynamical simulations. 
Using $\mathcal{V}$, we define the $\chi^2$ as
\begin{equation}
\label{eq:chi2}
\chi^2 = \mathcal{V} \mathcal{C}^{-1} \mathcal{V}^{T} ,
\end{equation}
where $\mathcal{C}$ is the covariance matrix. 

We define $\chi_{\rm{null}}^2$ by setting $A_{\rm{tSZ}}=0$.  % to quantify the level at which we reject the \textit{null} hypothesis of having no correlation between lensing and tSZ signals. 
In addition, $\chi_{\rm{min}}^2$ is found by minimizing Eq.~(\ref{eq:chi2}) with respect to $A_{\rm{tSZ}}$:

\begin{eqnarray}
%\begin{cases} 
                       \chi_{\rm{null}}^2 &:&   A_{\rm{tSZ}}=0  ,  \\
                       \chi_{\rm{min}}^2 &:&  A_{\rm{tSZ, min}} .
%\end{cases}
\label{eq:bin_operator}
\end{eqnarray}
In other words, $\chi_{\rm{min}}^2$ quantifies the goodness of fit between the measurements and our model prediction after marginalizing over $A_{\rm{tSZ}}$. 

Table \ref{tab:chi2} summarizes the $\chi_{\rm{null}}^2$ values from the measurements before and after including the sampling variance contribution. The values are quoted for individual estimators  as well as when they are combined. The $\chi_{\rm{null}}^2$ is always higher for $y-\gamma_t$ estimator, demonstrating that it is a better estimator for our cross-correlation analysis. It also improves when we combine the estimators but we should consider that $\chi_{\rm{null}}^2$ increases at the expense of adding extra degrees of freedom. Namely, we have 8 angular bins for each estimators and combining the two, there are 16 degrees of freedom which introduces a redundancy due to the correlation between the two estimators so that $\chi_{\rm{null}}^2$ does not increase by a factor of 2.

%%%%%%%%%%%%%%%%%%%%%%%%%%%%%%%
\begin{table}
\centering
\begin{tabular}{c|c|c|c|c|c|c|c|c|c}
\hline 
Estimator    & DoF &  $\chi_{\rm{null}}^2$, Stat. err. only  & $\chi_{\rm{null}}^2$, Adjusted \\
\hline
\hline
  $\xi^{y-\kappa}$        &   8  & 193.5 & 41.6 \\

  $\chi^{y-\gamma_t}$   &   8  & 307.4  & 65.5\\

  combined               &   16  & 328.7& 71.4\\
  \hline
$C_\ell^{y-\kappa}$&5                    &156.4                &60.1                 \\
    \hline
  \hline
  \end{tabular}
  \caption{A summary of the $\chi_{\rm{null}}^2$ values before and after including the sampling variance contribution according to the adjustment procedure of Section \ref{sec:error-inflate}. There are 8 angular bins, or degrees of freedom (DoF), at which the individual estimators are computed. Combining the estimators increases the DoF accordingly.}
\label{tab:chi2}
\end{table}
%%%%%%%%%%%%%%%%%%%%%%%%%%%%%%%

Finally, we define the SNR as follows. 
We wish to quantify how strongly we can reject the null hypothesis $H_0$, that no correlation exists between lensing and tSZ, in favour of the alternative hypothesis $H_1$, that the cross-correlation is well described by our fiducial model up to a scaling by the cross-correlation bias $A_\mathrm{tSZ}$. To this end, we employ a likelihood ratio method. The deviance $\mathcal{D}$ is given by the logarithm of the likelihood ratio between $H_0$ and $H_1$:
\begin{equation}
\mathcal{D} = -2\log\frac{\mathcal{L}(\vec d | H_0)}{\mathcal{L}(\vec d| H_1)}.
\end{equation}
For Gaussian likelihoods, the deviance can then be written as
\begin{equation}
\mathcal{D} = \chi^2_\mathrm{null} - \chi^2_\mathrm{min} \ .
\end{equation}
If $H_1$ can be characterised by a single, linear parameter, $\mathcal{D}$ is distributed as $\chi^2$ with one degree of freedom \citet{Williams2001}. The significance in units of standard deviations $\sigma$ of the rejection of the null hypothesis, i.e., the significance of detection, is therefore given by:
\begin{equation}
\rm{SNR} = \sqrt{\chi_{\rm{null}}^2 - \chi_{\rm{min}}^2}.
\label{eq:SNR}
\end{equation}

Table \ref{tab:chi2-A} summarizes the significance analysis of our measurements. We show %the $\chi_{null}^2$ values, as well as 
the SNR and best-fit amplitude $A_{\rm{tSZ}}$, for the theoretical halo model predictions with \textit{WMAP}-7yr and \textit{Planck} cosmologies. The results are presented for each estimator independently as well as for their combination. Note that all the values in Table \ref{tab:chi2-A} are adjusted to account for the sampling variance, as described in Section \ref{sec:error-inflate}. To estimate the combined covariance matrix, we place the covariance for individual estimators as block diagonal elements of combined matrix and compute the off-diagonal blocks (the covariance between the two estimators). 

The predictions from \textit{WMAP}-7yr cosmology are relatively favored in our analysis, which is consistent with the results of previous studies (e.g. \citealt{McCarthy2014, CFHT3}). We, however, find similar SNR values from both cosmologies because the effect of the different cosmologies on the halo model prediction can be largely accounted for by an overall rescaling ($A_{\rm{tSZ}}$). After rescaling, the remaining minor differences are due to the shape of the cross-correlation signal so that the SNR depends only weakly on the cosmology. 

We obtain a 13.2$\sigma$ and 16.7$\sigma$ from $y-\kappa$ and $y-\gamma_t$ estimators, respectively, when we only consider the statistical noise in the covariance matrix (before the adjustment prescription of Section \ref{sec:error-inflate}) \footnote{Note that we are quoting the detection levels from a statistical noise-only covariance matrix so that we can compare to the previous literature, including the results of \citet{CFHT1}, where a similar approach is taken in the construction of the covariance matrix (i.e. only statistical noise is considered).}. 

The 13.2$\sigma$ significance from $y-\kappa$ estimator should be compared to the $\sim$6$\sigma$ detection from the same estimator in \citet{CFHT1} where CFHTLenS data are used instead.  As expected, RCSLenS yields an improvement in the SNR and $y-\gamma_t$ improves it further. Including sampling variance in the covariance matrix decreases the detection significance from RCSLenS data to 6.1$\sigma$ and 7.7$\sigma$ for the $y-\kappa$ and $y-\gamma_t$ estimators, respectively.

%%%%%%%%%%%%%%%%%%%%%%%%%%%%%%%
\begin{table*}
\centering
\begin{tabular}{c|c|c|c|c|c|c|c|c|c}
\hline 
    Estimator    & SNR, Stat. err. only &   SNR, Adjusted  &    $A_{\rm{tsz}},$ \textit{WMAP}-7yr  &     $A_{\rm{tsz}},$ \textit{Planck}  \\
\hline
\hline

  $\xi^{y-\kappa}$        &   13.2 & 6.1 & 1.14 $\pm$ 0.19 & 0.61 $\pm$ 0.10 \\

  $\xi^{y-\gamma_t}$   &  16.7                 &7.7                  & 1.22 $\pm$ 0.16 & 0.66 $\pm$ 0.08 \\

  combined           &  17.1                 &8.0                  & 1.23 $\pm$ 0.15 & 0.66 $\pm$ 0.08 \\
\hline
$C_\ell^{y-\kappa}$    &  11.8                 &7.3                  & 0.98 $\pm$ 0.13 & 0.54 $\pm$ 0.07 \\
  \hline
  \hline
  \end{tabular}
  \caption{A summary of the statistical analysis of the %configuration-space 
  cross-correlation measurements. For the configuration-space estimators, the results are shown for each estimator independently and when they are combined. SNR quantifies the significance of detection after a fit to model predictions (halo model). SNR values are shown before (SNR, Stat. err. only) and after (SNR, Adjusted) adjustment for sampling variance uncertainties according to the description of Section \ref{sec:error-inflate}, while $A_{\rm{tSZ}}$ values are quoted after the adjustment. The \textit{Planck} cosmology predicts higher amplitude than \textit{WMAP}-7yr cosmology so that overall, the \textit{WMAP}-7yr cosmology predictions are in better agreement with the measurements.
}

  \label{tab:chi2-A}
\end{table*}
%%%%%%%%%%%%%%%%%%%%%%%%%%%%%%%

We perform a similar analysis in Fourier space where the data vector is given by the pseudo-$C_\ell$s and the results are included in Table \ref{tab:chi2-A}. The SNR values are in agreement with the configurations-space analysis\footnote{Note that the Fourier-space analysis is performed by pipeline 3 in \citet{Harnois-Deraps2016}. Different pipelines give slightly different but consistent results.}. We see a similar trend as in the configuration-space analysis in that there is a better agreement with the \textit{WMAP}-7yr halo model predictions ($A_{\rm{tSZ}}$ is closer to 1) while the \textit{Planck} cosmology predicts a high amplitude.

Table \ref{tab:chi2-A} summarizes the predictions from the halo model framework with a fixed pressure profile for gas (UPP). In Section \ref{sec:implications}, we revisit this by comparing to predictions from hydrodynamical simulations where halos with different mass and at different redshifts have a variety of gas pressure profiles. We show that we find better agreement with models where AGN feedback is present in halos.

\subsection*{Impact of maximum angular separation}

The two configuration-space estimators we use probe different dynamical scales by definition. This means that as we include cross-correlations at larger angular scales, information is captured at a different rate by the two estimators. For a survey with limited sky coverage, combining the two estimators will therefore improve the SNR of the measurements. In the following, we quantify this improvement of the SNR.

\begin{figure}
\includegraphics[width=1.0\columnwidth]{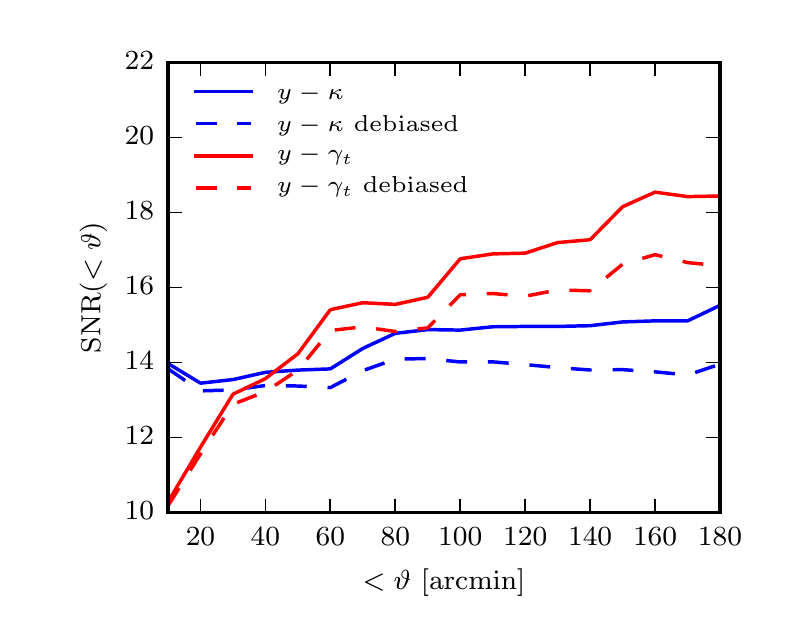}
%}}
\caption{The cumulative SNR %and reduced $\chi^2$ 
from RCSLenS as a function of the maximum angular separation for the two configuration-space estimators, with and without debiasing. %The plots are normalized to their values at 2 degree (120 arcmin) separation where the vertical lines are positioned.  
The slope of the lines is different for the two estimators due to the different information they capture as a function of angular separation.  For $y-\kappa$, most of the signal is in the first few bins making it a better candidate at small scales, while $y-\gamma_t$ has a larger slope and catches up quickly. Eventually, the two estimators approach a plateau where they contain the same amount of cross-correlation information.
}
\label{fig:nbin}
\end{figure}

In Fig.~\ref{fig:nbin}, we plot the cumulative SNR %and $\chi_{\nu}^2$ (reduced $\chi^2$ where $\nu$ is the number of degrees of freedom) 
values from RCSLenS as a function of the maximum angular separation for both estimators. In addition, we also include the same quantities when debiased using Eq.\ref{eq:Hartlap} to highlight the effect if increasing DoF on the debiasing factor.  Each angular bin is 10 arcmin wide and adding more bins means including cross-correlation at larger angular scales. 

We observe that the $y-\kappa$ measurement starts off with a higher SNR relative to  $y-\gamma_t$ at small angular separations. The SNR in $y-\kappa$ levels off very quickly with little information added above 1 degree separation. The shallow SNR slope of the $y-\kappa$ curve is partly due to the Gaussian smoothing kernel that is used in reconstructing the mass maps which spreads the signal within the width of the kernel. The $y-\gamma_t$ cross-correlation, on the other hand, has a higher rate of gain in SNR and catches up with the convergence rapidly. Eventually, the two estimators approach a plateau as the cross correlation signals drops to zero. At that point, both contain the same amount of cross-correlation information. 

Note that we limit ourselves to a maximum angular separation of 3 degrees in the RCSLenS measurements since the measurement is very noisy beyond that. Fig.~\ref{fig:nbin} indicates that the two estimators might not have converged to the limit where the information is saturated (a plateau in the SNR curve). Since each estimator captures different information up to 3 degrees, combining them improves the measurement significance (see Table \ref{tab:chi2-A}). With surveys like the Kilo-Degree Surveys (KiDS)  \citet{KiDS}, the Dark Energy Survey (DES) \citet{DES} and the Hyper Suprime-Cam Survey (HSC) \citet{HSC}, where the coverage area is larger, we will be able to go to larger angular separations where the information from our estimators is saturated. The signal at such large scales is primarily dependent on cosmology and quite independent of the details of the astrophysical processes inside halos (see \citealt{CFHT3} for more details). This could, in principle, provide a new probe of cosmology based on the cross-correlation of baryons and lensing on distinct scales and redshifts. 

\section{Implications for cosmology and astrophysics }
\label{sec:implications}

In Section \ref{sec:measurement} we compared our measurements to predictions from the theoretical halo model. The halo model approach, however, has limitations for the type of cross-correlation we are considering. For example, in our analysis we cross-correlate every sources in the sky that produces a tSZ signal with every source that produces a lensing signal. A fundamental assumption of the halo model, however, is that  all the mass in the Universe is in spherical halos, which is not an accurate description of the large-scale structure in the Universe. There are other structures such as filaments, walls or free flowing diffuse gas in the Universe, so that matching them with spherical halos could lead to biased inference of results (see \citealt{CFHT3}). Another shortcoming is that our halo model analysis considers a fixed pressure profile for diffuse gas, while it has been demonstrated in various studies that the UPP does not necessarily describe the gas around low-mass halos particularly well (e.g., \citealt{Battaglia2012,LeBrun2015}). 

Here we employ the cosmo-OWLS suite of cosmological hydrodynamical simulations (see Section \ref{sec:simulations}) which includes various simulations with different baryonic feedback models. The cosmo-OWLS suite provides a wide range of tSZ and lensing (convergence and shear) maps allowing us to study the impact of baryons on the cross-correlation signal. We follow the same steps as we did with the real data to extract the cross-correlation signal from the simulated maps.

\begin{figure}
%{\centering{
%\includegraphics[width=1.0\columnwidth]{plots/sims.pdf}
%\includegraphics[width=1.0\columnwidth]{plots/sims_AGN.pdf}
\includegraphics[width=1.0\columnwidth]{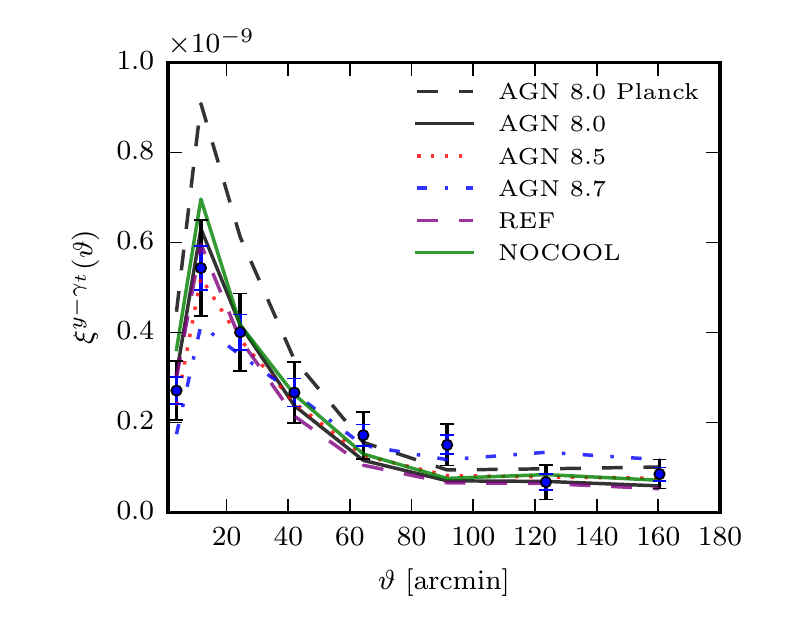}
%}}
\caption{Comparisons of our cross-correlation measurement from RCSLenS to predictions from hydrodynamical simulations. The larger (smaller) error bars represent uncertainties after (before) including our estimate of the sampling variance contribution (see Section \ref{sec:covariance}). Different baryon feedback models with \textit{WMAP}-7yr cosmology are shown for $y-\gamma_t$ estimator (we are not showing the plots for $y-\kappa$ as they are very similar). Baryon feedback has an impact on the cross-correlation signal at small scales.}
\label{fig:sims}
\end{figure}

Fig.~\ref{fig:sims} compares our measured configuration-space cross-correlation signal to those from simulations with different feedback models. %(we only plot the $y-\gamma_t$ estimator as the $y-\kappa$ predictions for different models are close and can not be distinguished). 
The plots are for the 5 baryon models using the \textit{WMAP}-7yr cosmology and the \textsc{AGN 8.0} model using the \textit{Planck} cosmolgy is also plotted for comparison. 
Note that baryon models make the largest difference at small scales due to mechanisms that change the density and temperature of the gas inside clusters. For the (non-physical) \textsc{NOCOOL} model, the gas can reach very high densities near the centre of dark matter halos and is very hot since there is no cooling mechanism in place.  This leads to a high tSZ and hence a high cross-correlation signal. After including the main baryonic processes in the simulation (e.g. radiative cooling, star formation, SN winds), we see that signal drops on small scales. Adding AGN feedback warms up the gas but also expels it to larger distances from the centre of halos. This explains why we see lower signal at small scales but higher signal at intermediate scales for the \textsc{AGN 8.7} model. Note that the scatter of the LoS signal varies for different models due to the details of the baryon processes so that, for example, the \textsc{AGN 8.7} model creates a larger sampling variance. The mean signal of the feedback models is also affected by the cosmological parameters at all scales. Adopting a \textit{Planck} cosmology produces a higher signal at all scales and for all models.  This is mainly due to the larger values of $\Omega_b$, $\Omega_m$, and particularly $\sigma_8$, in the \textit{Planck}~2013 cosmology compared to that of the \textit{WMAP}-7yr cosmology.

%**************************************
%**************************************

We summarize in Table \ref{tab:chi2-sims} our $\chi^2$ analysis for feedback model predictions relative to our measurements. We find that the data prefers a \textit{WMAP}-7yr cosmology to the \textit{Planck}~2013 cosmology {\it for all of the baryon feedback models}. This is worth stressing, given the very large differences between the models in terms of the hot gas properties that they predict \citep{LeBrun2014}, which bracket all of the main observed scaling relations of groups and clusters.  The best-fit models are underlined for both cosmologies. The \textsc{AGN 8.5} model fits the data best.

Our measurements are limited by the relatively low resolution of the \textit{Planck} tSZ maps. On small scales, our signal is diluted due to the convolution of the tSZ maps with the \textit{Planck} beam (FWHM = 9.5 arcmin) which makes it hard to discriminate the feedback models with our configuration-space measurements. This highlights that high-resolution tSZ measurements can be particularly useful to overcome this limitation and open up the opportunity to discriminate the feedback models from tSZ-lensing cross-correlations.

%%%%%%%%%%%%%%%%%%%%%%%%%%%%%%%
\begin{table*}
\centering
\begin{tabular}{c|c|c|c|c|c|c|c|c|c}
\hline 
&  Model  &   $\chi_{\rm{min}}^2$ , \textit{WMAP}-7yr  &    $A_{\rm{tsz}},$ \textit{WMAP}-7yr  &   $\chi_{\rm{min}}^2$, \textit{Planck}  &    $A_{\rm{tsz}},$ \textit{Planck}  \\
\hline
\hline
&\textsc{AGN 8.0} &  4.9 & 1.00 $\pm$ 0.13 &  \underline{4.9} & \underline{0.68 $\pm$ 0.09}\\
&\textsc{AGN 8.5} &  \underline{2.9} & \underline{1.10 $\pm$ 0.14} &  {5.4} &{ 0.74 $\pm$ 0.10} \\
Config. Space &\textsc{AGN 8.7} &  7.8 & 1.01 $\pm$ 0.13 &  7.6 & 0.72 $\pm$ 0.09 \\
&\textsc{NOCOOL}& 5.2  & 0.89 $\pm$ 0.11 &  5.3 & 0.62 $\pm$ 0.08 \\
&\textsc{REF} 		& 5.2  & 1.06 $\pm$ 0.14 & {5.3} & {0.74 $\pm$ 0.10}   \\
\hline
\hline
& \textsc{AGN 8.0} &  4.7 & 0.88 $\pm$ 0.12 &  4.9 & 0.61 $\pm$ 0.08\\
& \textsc{AGN 8.5} &  {2.4} & {1.05 $\pm$ 0.14} &  {2.8} & {0.71 $\pm$ 0.09} \\
Fourier Space& \textsc{AGN 8.7} &  \underline{0.7} & \underline{1.21 $\pm$ 0.16} &  \underline{0.9} & \underline{0.85 $\pm$ 0.11} \\
& \textsc{NOCOOL}&  6.6 & 0.78 $\pm$ 0.11 &  6.2 & 0.55 $\pm$ 0.11 \\
& \textsc{REF} 		&  5.5 & 0.92 $\pm$ 0.12 & 5.4 & 0.65 $\pm$ 0.09   \\
\hline
  \hline
  \end{tabular}
\caption{ Summary of $\chi_{\rm{min}}^2$ analysis of the cross-correlation measurements from hydrodynamical simulations. The error on the best-fit amplitudes are adjusted according to description of Section \ref{sec:error-inflate}. The best-fit models are underlined for both cosmologies. The \textit{WMAP}-7yr predictions fit data better for all baryon models. The \textsc{AGN 8.5} model is preferred by the configuration space measurements while the Fourier space measurements prefer the \textsc{AGN 8.7} model.
}
  \label{tab:chi2-sims}
\end{table*}
%%%%%%%%%%%%%%%%%%%%%%%%%%%%%%%

\begin{figure}
{\centering{
\includegraphics[width=1.0\columnwidth]{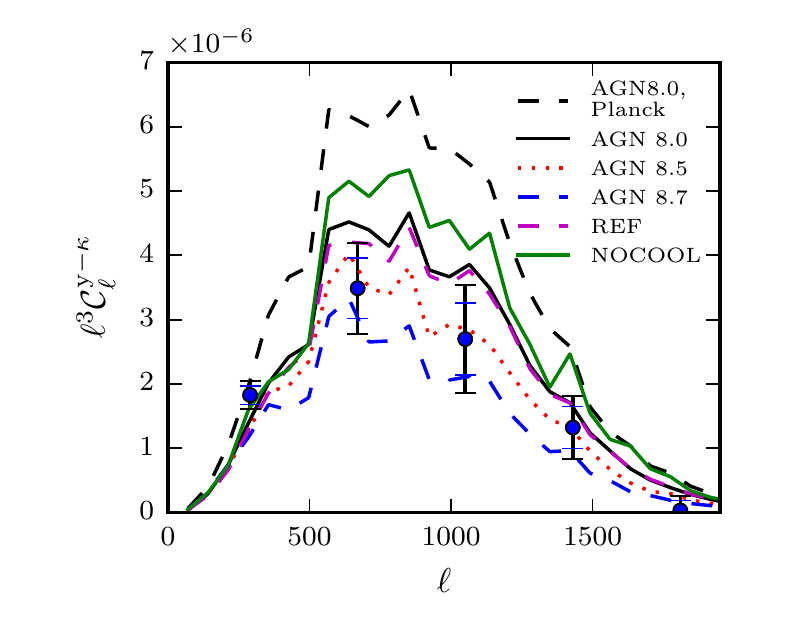}
}}
\caption{Same as Fig. \ref{fig:sims} for the Fourier-space estimator, $C_{\ell}^{y-\kappa}$. } 
\label{fig:sims-fourier}
\end{figure}

The feedback models could be better discriminated in Fourier-space through their power spectra. We therefore repeat our measurements on the simulation in the Fourier-space. We apply the same procedure described in Section \ref{sec:Fourier-space} to the simulated maps. Fig.~\ref{fig:sims-fourier} compares our power spectrum measurements to simulation predictions, and we have summarized the $\chi^2$ analysis in Table \ref{tab:chi2-sims}. 

%%%%%%%%%%%%%%%%%%%%%%%%%%%%%%%
%%%%%%%%%%%%%%%%%%%%%%%%%%%%%%%

Our Fourier-space analysis follows a similar trend as the configuration-space analysis. Namely, there is a general over-prediction of the amplitude which is worse for the textit{Planck} cosmology. For the best-fit \textit{WMAP}-7yr models the fitted amplitude is consistent with 1. Our estimator prefers the \textsc{AGN 8.7} model in all cases which is different than configuration-space estimator where the \textsc{AGN 8.5} was preferred\footnote{We tested that if large-scale correlations ($\vartheta > $ 100 arcmin), where uncertainties are large, are removed from the analysis, the \textsc{AGN 8.7} model is also equally preferred by the configuration-space estimators.}. There are a few reasons for these minor differences. For example, the binning schemes are different and give different weights to bins of angular separation.  Furthermore, the Fourier-based result can change by small amounts depending on precisely which pipeline from  \citet{Harnois-Deraps2016} is adopted.

When we fit the amplitude $A_{\rm{tSZ}}$ to obtain the $\chi_{\rm{min}}^2$ or SNR values, we are in fact factoring out the scaling of the model prediction and are left with the prediction of the shape of the cross-correlation signal. The shape depends on both the cosmological parameters (weakly) and details of the baryon model so that the values in Tables \ref{tab:chi2-A} and \ref{tab:chi2-sims} are a measure of how well the model shape matches the measurement. By comparing the $\chi_{\rm{min}}^2$ values from the two cosmologies in Table \ref{tab:chi2-sims}, the general conclusion of our analysis is that significant AGN feedback in baryon models is required to match the measurements well.

\section{Summary and Discussion}
\label{sec:summary}

We have performed cross-correlations of the public \textit{Planck} 2015 tSZ map with the public weak lensing shear data from the RCSLenS survey. We have demonstrated that such cross-correlation measurements between two independent data sets are free from contamination by residual systematics in each data set, allowing us to make an initial assessment of the implications of the measured cross-correlations for cosmology and ICM physics.

Our cross-correlations are performed at the map level where every object is contributing to the signal. In other words, this is not a stacking analysis where measurements are done around identified halos. Instead, we are probing all the structure in the Universe (halos, filaments etc) and the associated baryon distribution including diffuse gas in the intergalactic medium.

We performed our analysis using two configuration-space estimators, $\xi^{y-\gamma_t}$ and $\xi^{y-\kappa}$, and a Fourier-space analysis with $C_{\ell}^{y-\kappa}$ for completeness. Configuration-space estimators have the advantage that they are less affected by the details of map making processes (masks, appodization etc) and the analysis is straightforward. We showed that the estimators probe different dynamical scales so that combining them can improve the SNR of the measurement. 

Based only on the estimation of statistical uncertainties, the cross-correlation using RCSLenS data is detected with a significance of 13.2$\sigma$ and 16.7$\sigma$ from the $y-\kappa$ and $y-\gamma_{t}$ estimators, respectively. Including the uncertainties due to the sampling variance reduces the detection to 6.1$\sigma$ and 7.7$\sigma$, respectively. We demonstrated that RCSLenS data improves the SNR of the measurements significantly compared to previous studies where CFHTLenS data were used. 

The Fourier-space analysis, while requiring significant processing to account for masking effects, is more useful for probing the impact of physical effects at different scales. We work with $C_{\ell}^{y-\kappa}$ and test for consistency of the results with the configuration-space estimators.  We reach similar conclusions for the $C_{\ell}^{y-\kappa}$ measurement as the configuration-space counterpart as well as the same level of significance of detection.

The high level of detection compared to similar measurements in \citet{CFHT1} is due to two main improvements : the larger sky coverage offored by RCSLenS survey has suppressed the statistical uncertainties of the measured signal, and the final tSZ map provided by the \textit{Planck} team is also less noisy than that used in \citet{CFHT1}.  

We have compared our measurements against predictions from the halo model, which adopts the empirically-motivated `universal pressure profile' to describe the pressure of the hot gas associated with haloes.  Our significance analysis is based on a covariance matrix that includes statistical uncertainties only. While we are not able to construct a full covariance matrix which takes into account the uncertainties due to sampling variance, we are able to estimate its contribution. By scaling the measurement errors to account for sampling variance, we obtain realistic errors on the best-fit amplitude of the predictions. Predictions from a \textit{WMAP}-7yr best-fit cosmology match the data better than those based on the \textit{Planck} cosmology, in agreement with previous studies (e.g., \citealt{McCarthy2014}). 

Finally, we employed the cosmo-OWLS hydrodynamical simulations \citet{LeBrun2014}, using synthetic tSZ and weak lensing maps produced for a wide range of baryonic physics models in both the \textit{WMAP}-7yr and \textit{Planck} cosmologies.  In agreement with the findings of the halo model results, the comparison to the predictions of the simulations yields a preference for the \textit{WMAP}-7yr cosmology regardless of which feedback model is adopted.  This is noteworthy, given the vast differences in the models in terms of their predictions for the ICM properties of groups and clusters (which bracket the observed hot gas properties of local groups and clusters, see \citet{LeBrun2014}).  The detailed shape of the measured cross-correlations tend to prefer models that invoke significant feedback from AGN, consistent with what is found from the analysis of observed scaling relations, although there is still some degeneracy between the adopted cosmological parameters and the treatment of feedback physics.  Future high-resolution CMB experiments combined with large sky area from a galaxy survey can in principle break the degeneracy between feedback models and place tighter constrains on the model parameters. 

We highlighted the difficulties in estimating the covariance matrix for the type of cross-correlation measurement we consider in this paper. The large sampling variance in the cross-correlation signal, which originates mainly from the tSZ maps, requires access to more data or more hydrodynamical simulations to be accurately estimated. With the limited data we have, the covariance matrix that contained the contribution from sampling variance was noisy, making it impossible to perform a robust significance analysis of the measurements. This is an area where more work is required and we will pursue this in future work.

\section*{Acknowledgements}

% AH, TT, JHD, LvW, IGM, GH, YZM, HT, RCSLenS

We would like to thank the RCS2 team for conducting the survey, members of the RCSLenS team for producing the quality shear data, and the members of the OWLS team for their contributions to the development of the simulation code used here.

AH is supported by NSERC. TT is supported by the Swiss National Science Foundation (SNSF). JHD acknowledge support from the European Commission under a Marie-Sk{\l}odwoska-Curie European Fellowship (EU project 656869). IGM is supported by a STFC Advanced Fellowship at Liverpool JMU.  LVW, and GH are supported by NSERC and the Canadian Institute for Advanced Research. AC acknowledges support from the European Research Council under the EC FP7 grant number 240185. TE is supported by the Deutsche Forschungsgemeinschaft in the framework of the TR33 ‘The Dark Universe’. CH acknowledges support from the European Research Council under  grant number 647112.

This paper made use of the RCSLenS survey (\url{http://www.cadc-ccda.hia-iha.nrc-cnrc.gc.ca/en/community/rcslens/query.html}) and \textit{Planck} Legacy Archive (\url{http://archives.esac.esa.int/pla}) data.

%%%%%%%%%%%%%%%%%%%%%%%%%%%%%%%%%%%%%%%%%%%%%%%%%%

%%%%%%%%%%%%%%%%%%%% REFERENCES %%%%%%%%%%%%%%%%%%
%\input{References.tex}
%\input{References_sorted.tex}

% The best way to enter references is to use BibTeX:

%\input{RCSXy_V7.bbl}
\bibliographystyle{mnras}
\bibliography{mybib}

% Alternatively you could enter them by hand, like this:
% This method is tedious and prone to error if you have lots of references
%\begin{thebibliography}{99}
%\bibitem[\protect\citeauthoryear{Author}{2012}]{Author2012}
%Author A.~N., 2013, Journal of Improbable Astronomy, 1, 1
%\bibitem[\protect\citeauthoryear{Others}{2013}]{Others2013}
%Others S., 2012, Journal of Interesting Stuff, 17, 198
%\end{thebibliography}

%%%%%%%%%%%%%%%%%%%%%%%%%%%%%%%%%%%%%%%%%%%%%%%%%%

%%%%%%%%%%%%%%%%% APPENDICES %%%%%%%%%%%%%%%%%%%%%

\appendix

\section{Extra Considerations in $\kappa$-map reconstruction}
\label{sec:extras}

In the following, we describe the set of extra processing steps we have performed in our $\kappa$-map reconstruction pipeline to improve the SNR of our cross correlation measurements. These include the selection function applied to the lensing shear catalogue, adjustments in the reconstruction process, and proper masking of the contamination in the tSZ $y$ maps.

\subsection*{Magnitude Selection}
\label{sec:magcut}
One of the parameters that we can optimize to increase the SNR is the magnitude selection function  with which we select galaxies from the shear catalogue (and then make convergence maps from). This is not trivial as it is not obvious whether including faint sources with a noisy shear signal would improve our signal or not. To investigate this, we apply different magnitude cuts to our shear catalogue and compute the SNR of the correlation function measurement in each case.

\begin{figure*}
{\centering{
\includegraphics[width=1.0\columnwidth]{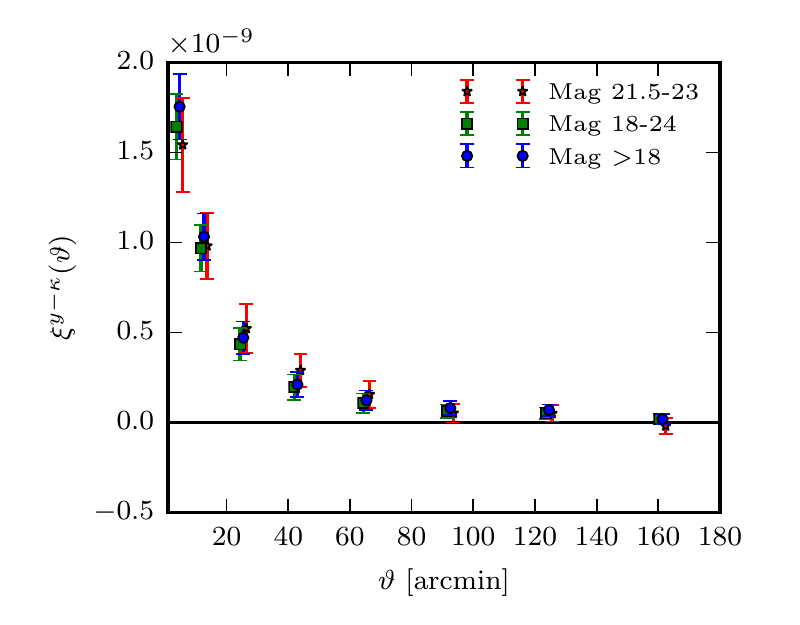}
\includegraphics[width=1.0\columnwidth]{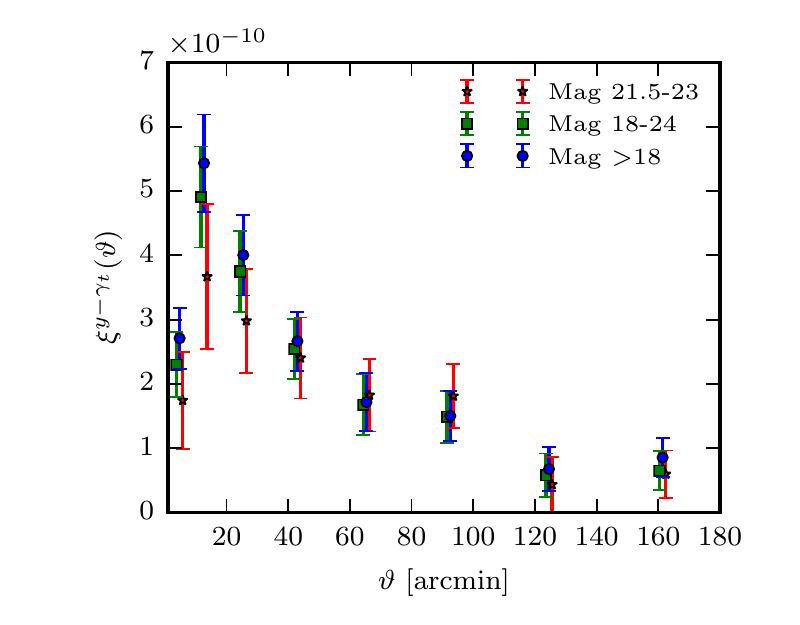}
}}
\caption{ Impact of different magnitude cuts on the $y-\kappa$ (left) and $y-\gamma_t$ (right)  cross-correlation signals. Including all the sources in the lensing surveys yields the highest SNR.}
\label{fig:cuts}
\end{figure*}

In Fig.~\ref{fig:cuts}, we compare the correlation functions from three different r-band magnitude cuts: $21-23.5$, $18-24$ and $> 18$ (all of the objects). We find  that the variations in the mean signal due to different magnitude cuts are relatively small. However, there is still considerable difference in the scatter around the mean signal which results in different SNR for the cuts. We consistently found that including all the objects (no cut) leads to a higher SNR.

%**************************************
%**************************************

\subsection*{Impact of smoothing}

Another factor that can change the SNR of the measurements is the smoothing kernel we apply to the lensing maps. Note that in our analysis, the resolution of the cross-correlation (the smallest angular separation)  is limited by the resolution of the tSZ maps which matches the observational beam scale from the \textit{Planck} satellite (FWHM$=9.5$ arcmin). On the other hand, lensing maps could have a much higher resolution and the interesting question is how the smoothing scale of the lensing maps affects the SNR.

As described before, the configuration-space $y-\gamma_t$ cross-correlation works at the catalogue level without any smoothing involved. However, in making the convergence maps, we apply a smoothing kernel as described in \citet{vanWaerbeke2013}. One has therefore the freedom to smooth the convergence maps with an arbitrary kernel. We consider three different smoothing scales and evaluate the SNR of the cross-correlations.

\begin{figure}
{\centering{
\includegraphics[width=1.0\columnwidth]{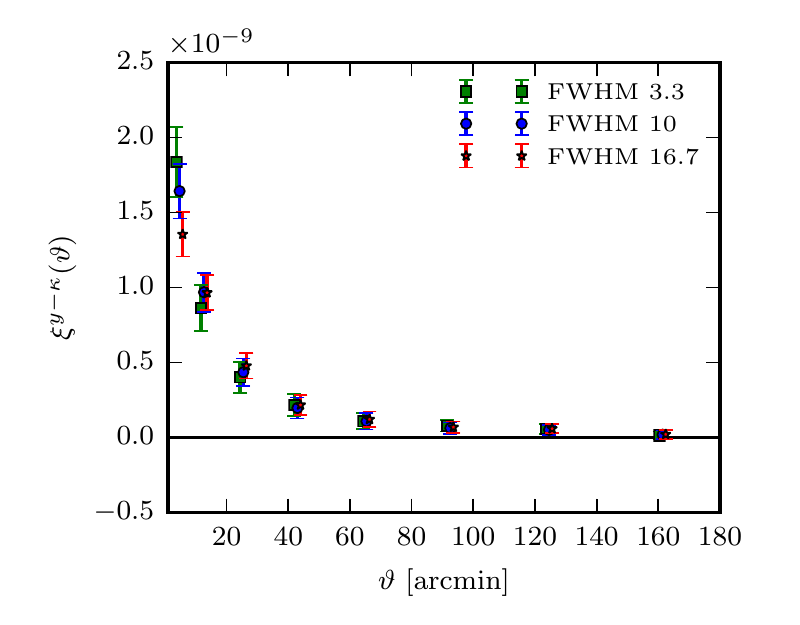}
}}
\caption{Impact of varying the smoothing of the convergence maps on the $y-\kappa$ cross-correlation signal. We apply three different smoothing scales of FWHM = 3.3, 10 and 16.7 arcmin during the mass map making process. While smaller scales result in a higher cross-correlation signal, the SNR is best for a smoothing scale of the same order as that of the $y$ maps (FWHM = 9.5 arcmin).  }
\label{fig:smoothing}
\end{figure}

Fig. \ref{fig:smoothing} demonstrates the impact of applying different smoothing scales (FWHM = 3.3, 10 and 16.5 arcmin) to the convergence maps used for the $y-\kappa$ cross-correlation. Note that while narrower smoothing kernels results in a higher cross-correlation amplitude, the uncertainties also increase and it lowers the SNR. %The complete $\chi^2$ analysis is shown in Table \ref{table:smooth_scale}.  
We concluded that smoothing the maps with roughly the same scale as the $y$ maps (FWHM $\approx 9.5$ arcmin) leads to the highest SNR.

%**************************************
%**************************************

\subsection*{Masks on the $y$ maps}

\begin{figure*}
{\centering{
\includegraphics[width=1.0\columnwidth]{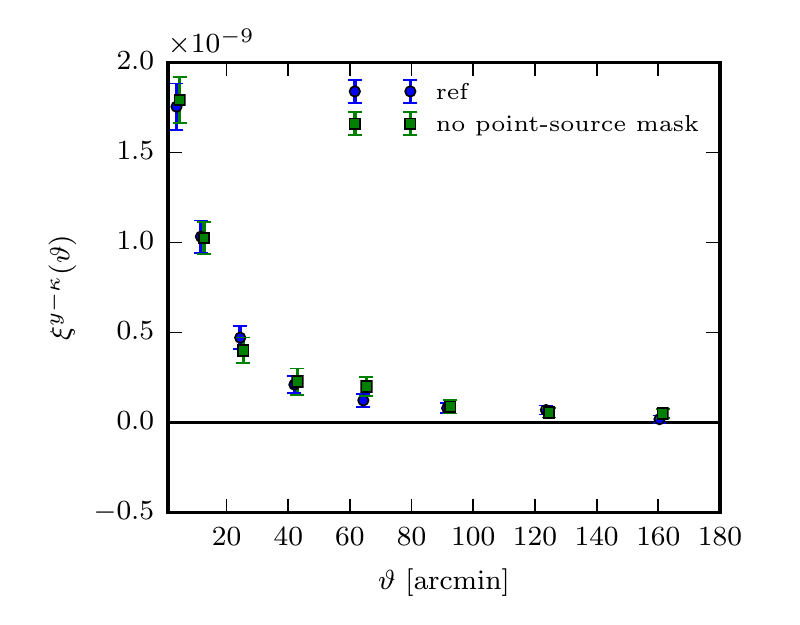}
\includegraphics[width=1.0\columnwidth]{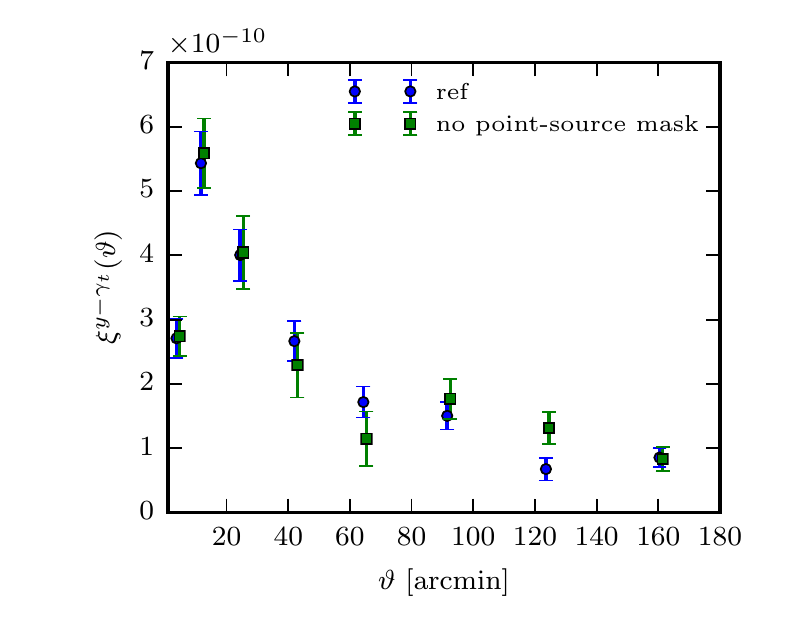}
}}
\caption{The impact of masking point sources in the $y$ map on the $y-\kappa$ (left) and $y-\gamma_t$ (right) cross correlation analysis. The measurements are fairly robust against such contamination. }
\label{fig:ps}
\end{figure*}

Since we work with tSZ maps provided by the \textit{Planck} collaboration, there is a minimal processing of the $y$ maps for our analysis. We apply the masks provided by the \textit{Planck} collaboration to remove point sources and galactic contamination. Note that the galactic mask does not significantly affect our measurements since all the RCSLenS fields are at high enough latitude. Cross-correlations are not sensitive to uncorrelated sources such as galactic diffuse emissions and point sources either. We have checked that our signal is robust against the masking of the point sources (see Fig.~\ref{fig:ps}).

%**************************************
%*******

%%%%%%%%%%%%%%%%%%%%%%%%%%%%
\section{Null tests and other effects}
\label{sec:null}

We have performed several consistency checks to verify our map reconstruction procedures and the robustness of the measurements. As mentioned before, an advantage of a cross-correlation analysis is that those sources of systematics that are unrelated to the measured signal will be suppressed in the measurement. This is particularly useful in the case of RCSLenS. As described below, there are residual systematics in the RCSLenS shear data (see \citet{RCSLenS} for details). It is therefore important to check if these systematics contaminate our cross-correlation. We start with a description of lensing $B$-mode residuals in RCSLenS data.

\subsection*{Lensing $B$-mode residuals}

In the absence of residual systematics, the scalar nature of the gravitational potential leads to a vanishing convergence $B$-mode signal.  As one of the important systematic checks in a weak lensing survey, one should investigate the level of the $B$-mode present in the constructed mass maps. This is a way to check that there is not any fake lensing signal due to the equipment or analysis deficiency since a true lensing signal is curl-free. 

To check for $B$-mode residuals in the RCSLenS data, we first create a new shear catalogue by rotating each galaxy in the original RCSLenS catalogue by $45$ degrees in the observation plane. This is equivalent to applying a transformation of shear components from $(\gamma_1, \gamma_2)$ to $(-\gamma_2, \gamma_1)$ \citet{Schneider98}.  We then follow our standard mass map making procedure to construct $B$-mode convergence maps, $\kappa_{B}^{\rm{obs}}$, from the new catalogue. Similar to the original maps, these maps are noisy and consists of the true underlying convergence, $\kappa_{B} $, and additional statistical noise, $\kappa_{ran}$:
\begin{equation}
\kappa_{B}^{\rm{obs}} = \kappa_{B} + \kappa_{ran} .
\end{equation}
It is, therefore, necessary to distinguish between the two components when searching for residual $B$-modes. 

To estimate $\kappa_{ran}$, we produce many ``noise" catalogue where this time the orientations of galaxies are randomly changed (these are essentially the same maps that are used to construct the covariance matrix as described in Section \ref{sec:covariance}). The constructed mass maps from these catalogue would only contain statistical noise. We estimate an average statistical noise auto-correlation function, $\bar{\xi}_{\kappa_{ran}}$, for each RCSLenS field by averaging over the auto-correlation function from the random mass map realizations of the field.  Finally, we estimate the residual $B$-mode signal in each of the RCSLenS fields by subtracting the statistical noise contribution computed for that field from the \textit{observed} auto-correlation function.

\begin{figure}
{\centering{
\includegraphics[width=1.0\columnwidth]{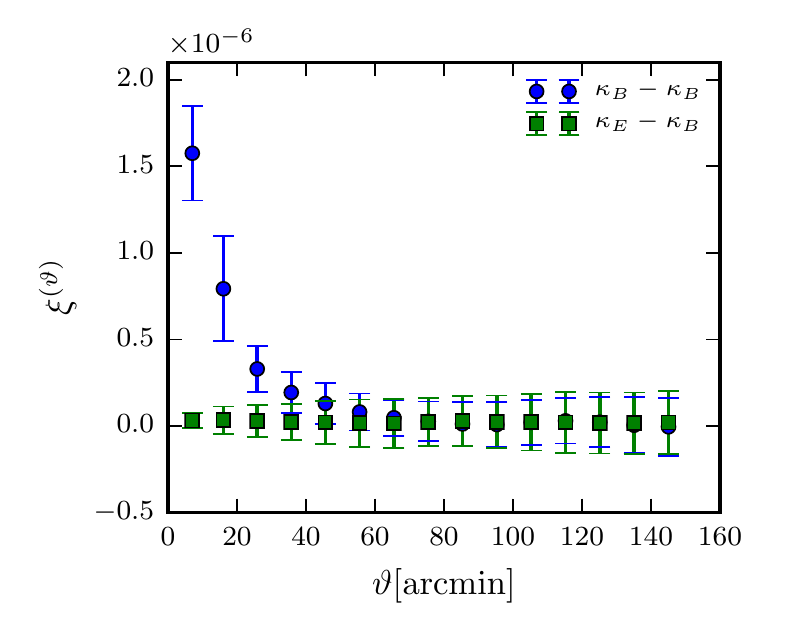}
}}
\caption{The stacked B-mode residual from the RCSLenS fields represented through the auto-correlation function, after subtracting the statistical noise contribution. The signal is not consistent with zero due to residual systematics in the shape measurements. The $\kappa_E - \kappa_B$ cross-correlation is also shown which is consistent with zero.}
\label{fig:Bmodes}
\end{figure}

Fig.~\ref{fig:Bmodes} shows the weighted average of the residual $B$-mode correlation function computed from the 14 RCSLenS fields after subtracting the contribution from statistical noise. The error bars represent the error on the mean value in each angular bin. Note that there is an excess of residuals $B$-mode at angular separations of $\leq 40$ arcmin. Independent analysis of  projected 3D shear power spectrum also confirms presence of excess residual $B$-mode signal at the corresponding scales \citep{RCSLenS}, consistent with our finding. The existence of such residual systematics could be problematic for our studies and needs to be checked as we describe in the following. 

In Fig.~\ref{fig:Bmodes}, we show $E$- and $B$-mode mass map cross-correlation from RCSLenS. The cross-correlation signal is consistent with zero which shows that any possible leakage of the systematic $B$-mode residuals to $E$-mode does not correlate with the true $E$-mode signal.

\subsection*{Residual tSZ-lensing systematic correlation}

%**************************************
\begin{figure*}
{\centering{
\includegraphics[width=1.0\columnwidth]{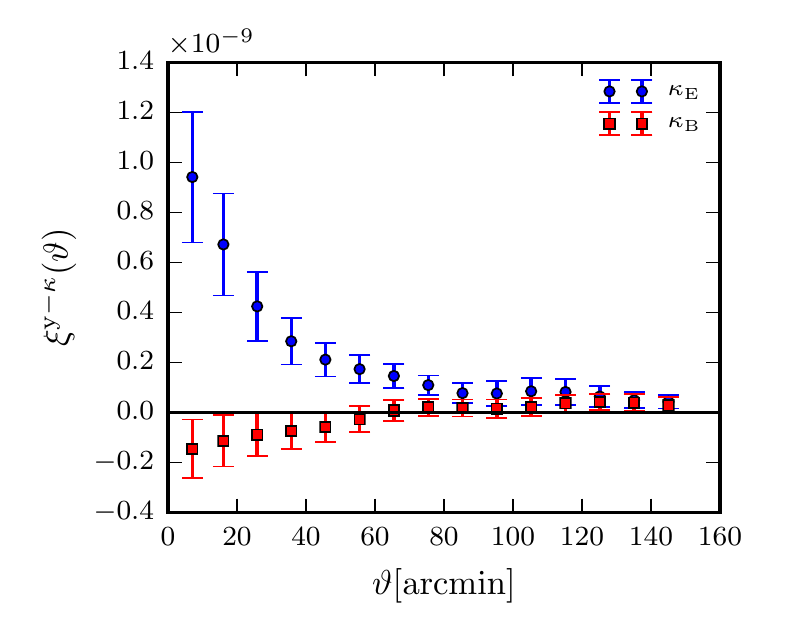}
\includegraphics[width=1.0\columnwidth]{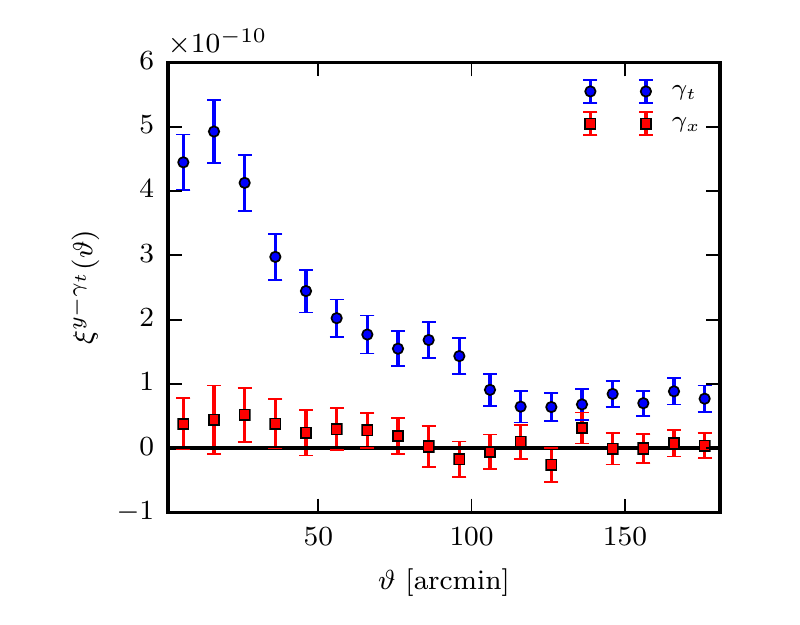}
}}
\caption{ A summary of the null tests performed on the $y-\kappa$ (left) and $y-\gamma_t$ (right) estimators. The red squares show the B-mode $\kappa$ (right) and $\gamma_\times$ cross-correlations which are consistent with zero as expected (The blue circles show the E-mode $\kappa$ (right) and $\gamma_t$ cross-correlations for comparison). The null tests are validated in all cases confirming that the level of contaminating systematics are under control in the cross-correlation analysis.}
\label{fig:tSZXB}
\end{figure*}
%**************************************

In the following, we demonstrate that while there is significant $B$-mode signal in the RCSLenS data, the lensing-tSZ cross-correlation signal is not contaminated. This serves as a good example of how cross-correlating different probes can suppress significant systematic residuals and make it useful for further studies.

As the first step, we cross-correlate the $y$ maps with the random noise maps constructed for each RCSLenS field. We computed the mean cross-correlation and the error on the mean from these random noise maps. Consistency with zero insures that there is not any unexpected correlation between the $y$ signal in the absence of a true lensing signal and insures that the field masks do not create any artifacts.

As the next step, we correlate the $y$ maps with the constructed $\kappa_B$ maps. This cross-correlation should also be consistent with zero to ensure that there is no unexpected correlation between the tSZ signal and the systematic lensing $B$-mode. Note that we can perform a similar consistency check using the shear data instead. The analog to the $\kappa_B$ mode for shear is the cross (or radial) shear quantity, $\gamma_\times$, defined as: 
\begin{equation}
\gamma_\times (\theta) = - \gamma_1 \cos(2\phi) + \gamma_2 \sin(2\phi) ,
\end{equation}
which can be constructed by 45 degree rotation of galaxy orientation in the shear catalogue. With this estimator, we expect the $y-\gamma_\times$ cross correlations to be consistent with zero as another check of systematics in our measurements.

Fig.~\ref{fig:tSZXB} summarizes the null tests described above. In the left panel, cross-correlations of $y$ with reconstructed $\kappa_B$ maps are shown with the $y-\kappa_{E}$ curve over-plotted for comparison. Correlation with $\kappa_{B}$ maps slightly deviates from zero at smaller scales due to residual systematics but is still insignificant considering the high level of bin-to-bin correlation. In the right panel, we show cross-correlation with $\gamma_\times$ with the $y-\gamma_{t}$ curve over-plotted for comparison. We don't see any inconsistency in the $y-\gamma_\times$ correlation. Both estimators are also perfectly consistent with zero when cross-correlated with random maps.

We therefore conclude that the systematic residuals are well under control and do not affect our measurements.

%**************************************
%**************************************

\bsp

\label{lastpage}

\end{document}